\documentclass[prl,reprint,longbibliography,superscriptaddress]{revtex4-1}
\usepackage{natbib,units}

\usepackage{amsmath,amssymb}
\usepackage{bbold}
\usepackage{dsfont}
\usepackage{dcolumn}
\usepackage{tikz,hyperref}
\usepackage[utf8]{inputenc}
\usepackage{verbatim}
\usepackage{mathtools}
\usepackage{animate,media9}
\usepackage{slashed}
\usepackage{braket}
\usepackage{subfigure}
\usepackage{slashed} 
\usepackage{hyperref}

\usepackage{graphicx}
\usepackage{bm}

\definecolor{lime}{HTML}{A6CE39}
\DeclareRobustCommand{\orcidicon}{%
	\begin{tikzpicture}
	\draw[lime, fill=lime] (0,0) 
	circle [radius=0.16] 
	node[white] {{\fontfamily{qag}\selectfont \tiny ID}};
	\draw[white, fill=white] (-0.0625,0.095) 
	circle [radius=0.007];
	\end{tikzpicture}
	\hspace{-2mm}
}

\foreach \x in {A, ..., Z}{%
	\expandafter\xdef\csname orcid\x\endcsname{\noexpand\href{https://orcid.org/\csname orcidauthor\x\endcsname}{\noexpand\orcidicon}}
}

\newcommand{\apjl}{APJL}
\newcommand{\mnras}{MNRAS}

\newcommand{\physrep}{Physics Reports}

\begin{document}

\title{A graph model for the clustering of dark matter halos}

\author{Daneng Yang \orcidA{}}
\email{danengy@ucr.edu}
\affiliation{Department of Physics and Astronomy, University of California, Riverside, California 92521, USA}

\author{Hai-Bo Yu \orcidB{}}
\email{haiboyu@ucr.edu}
\affiliation{Department of Physics and Astronomy, University of California, Riverside, California 92521, USA}

\date{\today}

\begin{abstract}

We use network theory to study topological features in the hierarchical clustering of dark matter halos. We use public halo catalogs from cosmological N-body simulations and construct tree graphs that connect halos within main halo systems. Our analysis shows these graphs exhibit a power-law degree distribution with an exponent of $-2$, and possess scale-free and self-similar properties according to the criteria of graph metrics. We propose a random graph model with preferential attachment kernels, which effectively incorporate the effects of minor mergers, major mergers, and tidal stripping. The model reproduces the structural, topological properties of simulated halo systems, providing a new way of modeling complex gravitational dynamics of structure formation.

\end{abstract}

\maketitle

{\noindent\bf Introduction.} Complex systems commonly contain many interacting components, and the connectivity of their components can be described using a graph~\cite{Albert2002,Boccaletti2006}. When graph nodes and edges are associated with physical components and interactions, respectively, a graph is also called a network. The number of edges connected to a node is its degree. Growing random network models have been constructed to study different systems, e.g., the world-wide-web, e-mail, social, protein, and metabolic networks~\cite{jeong2000large,albert2000topology,pastor-satorras_vespignani_2004,PhysRevLett.87.258701,pastor2004topology,PhysRevE.67.056104,broder2000graph,2001Natur.411...41J,Grossman95ona,watts1998collective,amaral2000classes,2002PhRvE..66c5103E}. These networks usually exhibit power-law degree distributions as $P(n)\propto n^{-\nu}$, where $P(n)$ is the relative frequency of the nodes with degree $n$, and $\nu$ is a constant with its value typically in the range $2<\nu<3$~\cite{Boccaletti2006}. This feature is known as the scale-free property.

Ref.~\cite{Barabasi1999} shows that a mechanism based on preferential attachment can explain the power-law degree distribution. In this mechanism, one starts with a connected graph and attaches a new node to the existing ones once at a time. 
The probability for a new node to be attached to the $i^{\rm th}$ existing node is ${A(n_i)}/{\sum_j A(n_j)}$, where $A(n_i)$ is an attachment kernel and $n_i$ is the degree of the $i^{\rm th}$ node. The summation in the denominator goes over all existing nodes~\cite{Albert2002,Boccaletti2006}.
If $A(n_i)$ increases with $n_i$, the growth of a graph naturally leads to a preference in the attachment, resulting in the effect that the rich get richer~\cite{RevModPhys.74.47,BOCCALETTI2006175,2015EPJB...88..234H,Price1976AGT,Barabasi1999,PhysRevLett.85.4629,PhysRevLett.85.4633}. When the attachment kernel is linear $A_i = n_i$, we have the Barab\'asi-Albert model that produces a distribution of $P(n)\propto n^{-3}$ at large $n$~\cite{Barabasi1999}. if the attachment kernel is nonlinear, the power-law feature of the degree distribution can be transient~\cite{Krapivsky2000,Krapivsky2001,2000cond.mat.11029B}.

In this work, we study the hierarchical clustering of dark matter halos using graphs. We will analyze state-of-the-art cosmological N-body simulations of structure formation from the FIRE2~\cite{2015MNRAS.450...53H,2018MNRAS.480..800H,Wetzel:2022man} and IllustrisTNG~\cite{Nelson:2018uso,Nelson:2019jkf,Pillepich:2019bmb} projects and construct graphs for individual main halo systems by connecting their associated subhalos. These graphs exhibit a power-law degree distribution with an exponent of $-2$. We will develop a random graph model to reproduce topological, structural properties of these graphs. In this model, the attachment kernel is linear for effectively modeling minor mergers in structure formation, while it approaches two asymptotic regimes for incorporating major mergers and tidal stripping. We will also use the metric proposed in~\cite{Li2005} and show the cosmic structure formation is scale-free {\it and} self-similar, and our graph model can reproduce this important feature. In the Appendices, we provide additional materials about the analysis for this work.

\begin{figure}
   \centering
  \includegraphics[height=4.2cm]{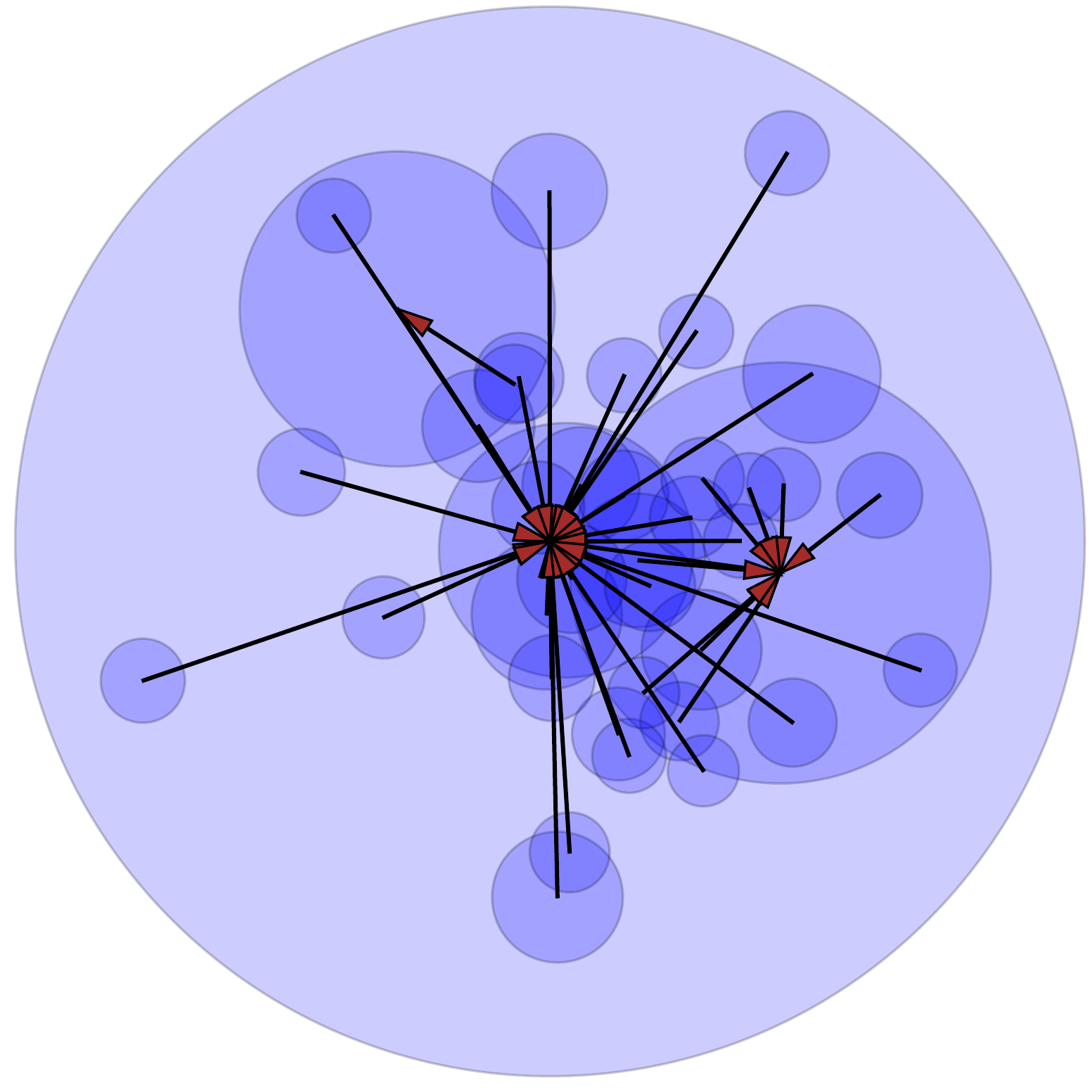}~~~
  \includegraphics[height=4.2cm]{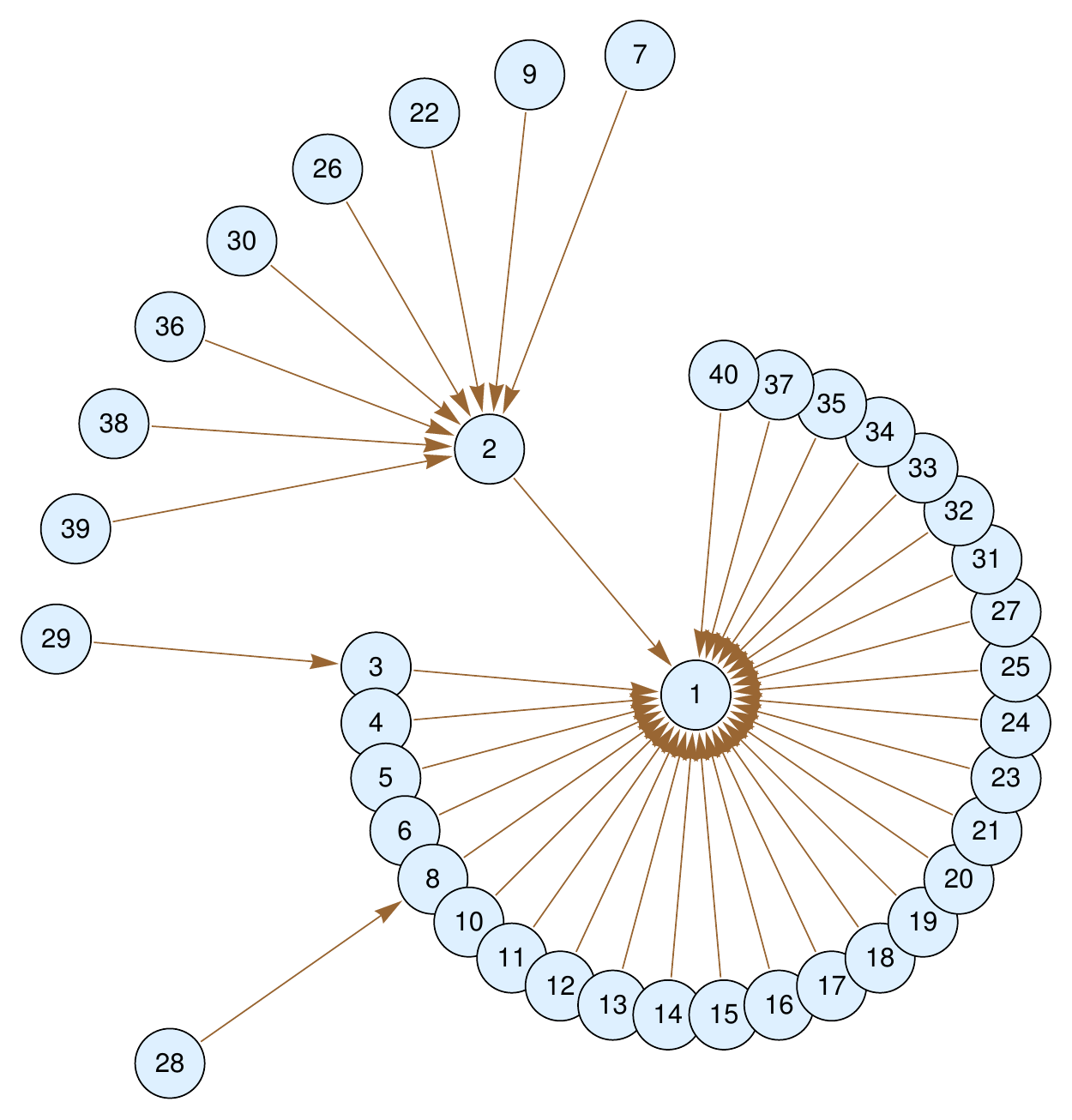} 
  \caption{   \label{fig:cartoon} 
  An example graph constructed for a Milky Way analog named m12r from the FIRE2 cosmological simulation~\cite{2020MNRAS.491.1471S,Wetzel:2022man}. {\it Left:} $40$ most massive dark matter halos of m12r. Each circle represents a dark matter halo and its size denotes the halo's virial radius. An arrowed link is created between a subhalo to its least massive host halo. {\it Right:} A tree graph for modeling the hierarchical clustering of the $40$ halos. A node represents a halo and an edge represents a connection of a subhalo to its least massive host halo. }
\end{figure}

{\noindent\bf Graphs for the clustering of subhalos.} In the standard model of cosmology, the cosmic structure forms hierarchically, i.e., small dark matter halos, a gravitationally self-bound system, form first, and they merge to form larger and more massive halos, see, e.g.,~\cite{Gao:2004au,Springel:2005nw,Springel:2008cc,Jiang2016}. During this process, most of the merging halos ``dissolve'', virialize, and become the smooth component of their host halos. However, some of them can survive from tidal disruption during the mergers and become subhalos. Fig.~\ref{fig:cartoon} (left) illustrates a Milky Way analog, named m12r, from the public data release of the FIRE2 simulation~\cite{Wetzel:2022man,2020MNRAS.491.1471S}. We include a main halo and its $39$ most massive subhalos. Here, we consider a main halo to be isolated and its mass and virial radius are largest in the system~\footnote{The virial radius of a halo is determined as the radius inside of which the average density is some constant times the matter density of the universe. The FIRE2 simulation considers a constant of $200$ for halos at redshift $z=0$. The TNG50 halo catalog has multiple choices and we take the definition introduced in~\cite{1998ApJ...495...80B}. The small difference in determining the virial radius does not affect our results.}. A subhalo of a main halo is identified if the distance between two halo centers is less than the virial radius of the latter, and we do not require the subhalo to be entirely enclosed within the virial radius of the main halo. Since structure formation is hierarchical, a subhalo can contain high-order subhalos, and itself can be directly hosted by another subhalo. For a given subhalo, we apply similar distance criteria to identify its higher-order subhalos.

To explore the hierarchical nature of structure formation, for each subhalo, we identify its least massive host halo, which is not necessarily the main halo, and create a link between the two; see the red arrows in Fig.~\ref{fig:cartoon} (left). We obtain a connected and directed tree graph that has a total number of $40=39$ (subhalos)+$1$ (main halo) nodes, and each of the nodes represents a dark matter halo; see Fig.~\ref{fig:cartoon} (right). For a constructed graph, the in-degree of the $i^{\rm th}$ node $N^i_{\rm id}$ is the number of arrows pointing to the node, corresponding to the number of subhalos that are {\it directly} hosted by the $i^{\rm th}$ halo. For m12r shown in Fig.~\ref{fig:cartoon} (right), there are four nodes that have none zero $N_{\rm id}$ values: $\{N^{1}_{\rm id},N^{2}_{\rm id}, N^{3}_{\rm id},N^{8}_{\rm id}\}=\{29,8,1,1\}$. We clearly see the hierarchical structure of the simulated system. For example, halo ``28" is the subhalo of halo ``8", which is in turn a subhalo of the main halo labeled as ``1". Note $N_{\rm sub}=\sum_i N^{i}_{\rm id}$, where $N_{\rm sub}$ is the number of subhalos of all orders up to the mass cut.

\begin{figure}[!htp]
  \centering
  \includegraphics[height=8.2cm]{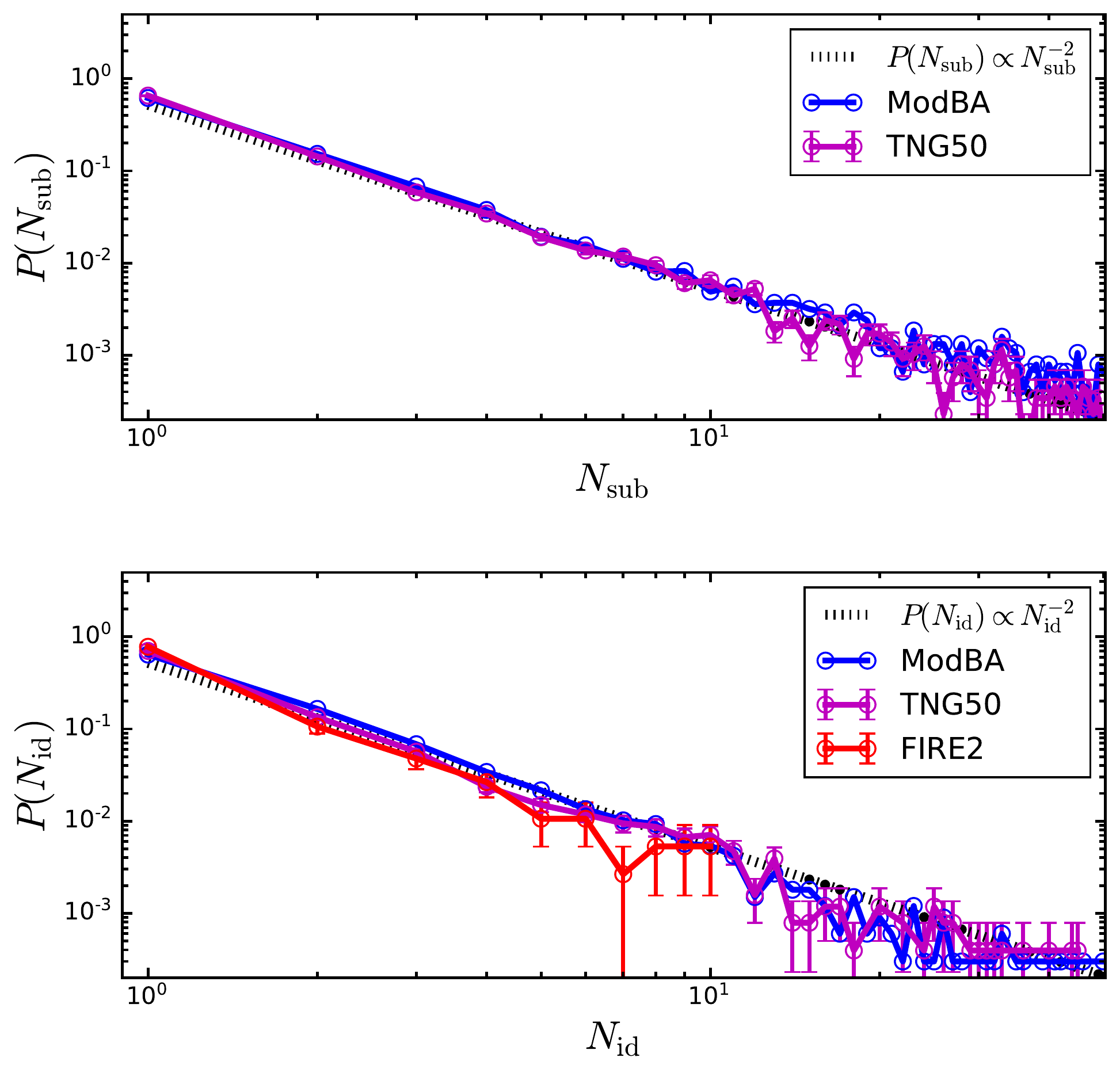}
  \caption{\label{fig:graph0} {\it Top:} the fraction of the main halos that have $N_{\rm sub}$ subhalos of all orders for graphs constructed from the TNG50 cosmological simulation (solid magenta) and our graph model, ModBA (solid blue), based on the attachment kernel Eq.~\ref{eq:kernel1}. {\it Bottom}: the in-degree distribution of the nodes for graphs from the TNG50 (solid magenta) and FIRE2 (solid red) simulations, as well as the ModBA model using the kernel Eq.~\ref{eq:kernel2} (solid blue). In both panels, a power-law scaling relation with an exponent of $-2$ is included for comparison (dotted black).
}
\end{figure}

{\noindent\bf The power-law degree distribution.} To construct a large sample of graphs and perform statistical analysis, we take public halo catalogs from two independent simulation projects. The FIRE2 simulation of galaxy formation is part of the Feedback In Realistic Environments (FIRE) project, generated using the Gizmo code~\cite{2015MNRAS.450...53H} and the FIRE2 physics model~\cite{2018MNRAS.480..800H}. 
We use the $11$ hydrodynamic zoom-in simulations in the FIRE2 data release with main halos masses $\sim 10^{12}~\rm M_{\odot}$~\cite{Wetzel:2022man,2019MNRAS.490.4447W,2018MNRAS.478..906C,2018MNRAS.473.1930E,2018MNRAS.480..800H,2019MNRAS.487.1380G,2017MNRAS.471.1709G,2016ApJ...827L..23W,2020MNRAS.491.1471S,2019MNRAS.489.4574G,2017MNRAS.472L.109A}. For each of the FIRE2 main halos, the resolution is high enough to resolve subhalos with masses larger $10^7~{\rm M_\odot}$, as the mass of simulations particles at the finest level can reach $3.5\times10^4~{\rm M_\odot}$. On average, there are $N_{\rm sub}\sim800$ resolved subhalos for each main halo. In our analysis, we further impose a mass cut and take the $600$ most massive ones, including the main halo, and have checked that our main results do not change if we vary the cut. 

The IllustrisTNG project is a suite of cosmological galaxy formation simulations~\cite{Pillepich:2017jle,2017MNRAS.465.3291W}. We use halo catalogs from the TNG-50-1-Dark (TNG50) simulation, which is a dark matter-only simulation and has the highest resolution~\cite{Nelson:2018uso,Nelson:2019jkf,Pillepich:2019bmb}. We consider (sub)halos of masses heavier than $5.4\times 10^7~{\rm M_\odot}$, i.e., at least a factor of $100$ larger than the mass of simulation particles. With this selection, we have about $8800$ TNG50 main halos with masses in the range $5.4\times10^{7}\textup{--}2.1\times 10^{14}~{\rm M_\odot}$, which contain at least one subhalo. The TNG50 and FIRE2 simulations use different programs to identify halos. However, the difference only affects some halos around the resolution limit and has negligible effects on our main results.

Fig.~\ref{fig:graph0} (top) shows the distribution of the subhalo number for the TNG50 main halos at redshift $z=0$ (solid magenta), where $P(N_{\rm sub})$ is the fraction of the main halos that have $N_{\rm sub}$ subhalos of all orders. The probability of having a high number of subhalos decreases and the trend follows a power-law scaling relation as $P(N_{\rm sub})\propto N^{-2}_{\rm sub}$ (dotted black). Since the $11$ FIRE2 main halos have similar masses, they are not suitable for studying the $N_{\rm sub}$ distribution. From the perspective of structure formation theory, the scaling relation $P(N_{\rm sub})\propto N^{-2}_{\rm sub}$ can be attributed to the scale invariance of the matter power spectrum. In this case, the Press-Schechter formalism~\cite{1974ApJ...187..425P} predicts the mass function as $d n/d M \propto 1/M^2$ on the low mass limit for isolated main halos, where $n$ is the number of halos and $M$ is the halo mass. Thus we have $d n/d N_{\rm sub} \propto 1/N^2_{\rm sub}$ as $N_{\rm sub}\propto M$ approximately~\cite{Zentner:2004dq}. 

Fig.~\ref{fig:graph0} (bottom) shows the in-degree distribution $P(N_{\rm id})$ for the FIRE2 (solid red) and TNG50 (solid magenta) simulations. We evaluate $P(N_{\rm id})$ in the following way. Consider the graph shown in Fig.~\ref{fig:cartoon} as an example:  among the four nodes that has non-zero in-degree, two of them has in-degree $1$, and hence $P(N_{\rm id}=1)=2/4=0.5$; one has $29$, $P(N_{\rm id}=29)=1/4=0.25$; one has $8$, $P(N_{\rm id}=8)=1/4=0.25$. Following the procedure, we construct graphs with $600$ nodes (halos) and obtain the averaged $P(N_{\rm id})$ distributions over $11$ FIRE2 and $67$ TNG50 systems, respectively. We again see the $N_{\rm id}$ distribution exhibits a power law of $P(N_{\rm id})\propto N_{\rm id}^{-2}$ (dotted black) for both simulations, although they have different implementations, initial conditions and resolutions.

The results shown Fig.~\ref{fig:graph0} are robust to baryonic feedback associated with galaxy formation. The FIRE2 halos are from hydrodynamic simulations, while the TNG50 halos we analyzed are dark-matter-only. Nevertheless, their $N_{\rm sub}$ and $N_{\rm id}$ distributions are almost identical. We have further constructed graphs for the TNG50-1 simulation, which is the hydrodynamic version of TNG50-1-Dark, and found that the normalized subhalo number and in-degree distributions remain the same. 

Moreover, we have analyzed the halos from the Bolshoi dark matter simulation with Planck cosmology~\cite{Klypin:2014kpa}. Its box size is larger than the TNG50 one, and it contains more massive halos: $210$ main halos with masses in the range $10^{14}\textup{--}10^{15}~{\rm M_\odot}$, and four halos larger than $10^{15}~{\rm M_\odot}$. The resulting $N_{\rm sub}$ and $N_{\rm id}$ distributions from Bolshoi agree with those from FIRE2 and TNG50. Thus the distributions shown in Fig.~\ref{fig:graph0} are valid for main halo systems from dwarf to cluster scales, the full mass range.

The $N_{\rm sub}$ and $N_{\rm id}$ distributions provide complementary information about the hierarchical clustering of dark matter halos. For a given main, isolated halo, $N_{\rm sub}$ is the total number of subhalos of all orders. Thus $P(N_{\rm sub})$ measures the probability of having a graph with the number of nodes $N=N_{\rm sub}+1$ in the structure formation. In contrast, the in-degree distribution $P(N_{\rm id})$ characterizes detailed structural, topological properties among the $N$ nodes {\it within} a graph. As we will discuss in the next section, two different attachment kernels are required to produce $P(N_{\rm sub})$ and $P(N_{\rm id})$ constructed from the N-body simulations.

{\noindent\bf A modified Barab\'asi-Albert model for dark matter halos.} The scaling relation shown in Fig.~\ref{fig:graph0} implies an increment in $N_{\rm sub}$ must be proportional to its current value, i.e., $\Delta N_{\rm sub}\propto N_{\rm sub}$ such that the power-law distribution is maintained. For minor mergers, where a low mass halo falls into a massive one, this halo accretion process is equivalent to an attachment with the kernel to be linear in $N_{\rm sub}$. However, as indicated in the original Barab\'asi-Albert model, for a linear kernel, the degree distribution follows a power law with an exponent of $-3$. For the simulated dark matter halos, both $N_{\rm sub}$ and $N_{\rm in}$ distributions have a power law with an exponent of $-2$. Thus the linear kernel, which mimics minor mergers in halo accretion, should be broken to some extent. 

We identify two physical processes that could break the linearity in two opposite limits. One is the major merger, i.e., two halos of similar mass coalesce into one, but without increasing the number of subhalos in the merged main halo. The other is tidal stripping, which removes subhalo mass in the outer region and redistributes it to the host halo. As a subhalo evolves in the tidal field of its host halo, it becomes lighter and smaller. Some higher-order subhalos residing in the outer region of the subhalo would be released, if not erased, resulting in a net enhancement in the attachment rate.  

With these considerations, we propose a modified Barab\'asi-Albert model (ModBA) to describe the clustering of dark matter halos. To reproduce the $N_{\rm sub}$ distribution, we consider the following attachment kernel
\begin{equation}
\label{eq:kernel1}
A_{j} = \frac{\alpha n_j^{\beta+1}}{ n_j + \alpha n_j^{\beta} }=\left\{
\begin{array}{ll} 
\alpha n^\beta_j ~~{\rm small}~~n_j\\
n_j~~{\rm large}~~ n_j
\end{array}
\right.
\end{equation}
where $n_j$ refers to the degree of the $j^{\rm th}$ node, $\alpha$ and $\beta$ are parameters to be determined through calibration with the N-body simulations. We see that $A_j$ asymptotes a linear kernel for large $n_j$, but it is suppressed for small $n_j$ if $\alpha\ll1$. This suppression may reflect the fact that the coalescing process during major mergers does not increase the number of subhalos, and this effect is most relevant for halos with low $N_{\rm sub}$. Taking $\alpha=0.05$ and $\beta=4.5$, we can produce the $N^{-2}_{\rm sub}$ distribution with constructed graphs, as shown in Fig.~\ref{fig:graph0} (top, solid blue). We have averaged over $8$ random graphs using the kernel Eq.~\ref{eq:kernel1} and each graph has $10^4$ nodes. We set $N_{\rm sub}$ to be the in-degree of the nodes.

The attachment kernel Eq.~\ref{eq:kernel1} is similar to the one proposed in~\cite{Krapivsky2000,Krapivsky2001}. They showed if one sets $A_1$ to be a constant $\alpha$, while keeping $A_j=n_j$ linear for $j\geq 2$, the constructed network will possess a central hub and has a power-law degree distribution for large $n$. The corresponding exponent is $-(3+\sqrt{1+8\alpha})/2$, which asymptotes $-2$ as $\alpha$ approaches zero. Thus in our model, the power-law distribution with the exponent $-2$ is realized naturally and it does not suffer from fine-tuning.

While $P(N_{\rm sub})$ measures the fraction of the main halos that have $N_{\rm sub}$ subhalos of all orders in the structure formation, $P(N_{\rm id})$ characterizes topological properties of a main halo system, i.e., how $N=N_{\rm sub}+1$ nodes are connected {\it within} a graph. The kernel in Eq.~\ref{eq:kernel1} is not well suitable for generating graphs for {\it individual} main halo systems, e.g., Fig.~\ref{fig:cartoon}, and their associated $P(N_{\rm id})$; see App.~\ref{app:comparison} for detailed discussion. Instead, we consider the following attachment kernel
\begin{equation}
\label{eq:kernel2}
A_{j} = n_j+\alpha n_j^{\gamma} =\left\{
\begin{array}{ll} 
n_j ~~{\rm small}~~n_j\\
\alpha n^\gamma_j~~{\rm large}~~ n_j
\end{array}
\right.
\end{equation}
where we choose $\alpha=0.05$ as before and $\gamma$ is a parameter depending on the total number of nodes (halos) $N$ in a graph. This kernel is linear for small $n_j$, while it amplifies the attachment rate for large $n_j$. This is consistent with the expectation that larger halos in the system acquire more subhalos released due to the effect of tidal stripping. 
We have calibrated $\gamma$ using the FIRE2 and TNG50 simulations and found that a relation of $\gamma={4.97}/{N^{0.117}}+{85.7}/{N^{1.44}}$ works for both.
We will discuss details of the calibration in the next section. 

In Fig.~\ref{fig:graph0} (bottom, solid blue), we show the $N_{\rm id}$ distribution, averaged over $100$ graphs randomly generated using the kernel Eq.~\ref{eq:kernel2} with $N=600$. It follows a power law with an exponent of $-2$, consistent with the $N_{\rm id}$ distribution from the simulated halos. For $N=600$, $\gamma\approx2.4$ and it does not become $1$ until $N\sim 10^6$. Thus the attachment kernel for low-degree nodes is still linear, while the one with a large degree receives an enhancement.

\begin{figure}[!htp]
\centering
  \includegraphics[width=8.2cm]{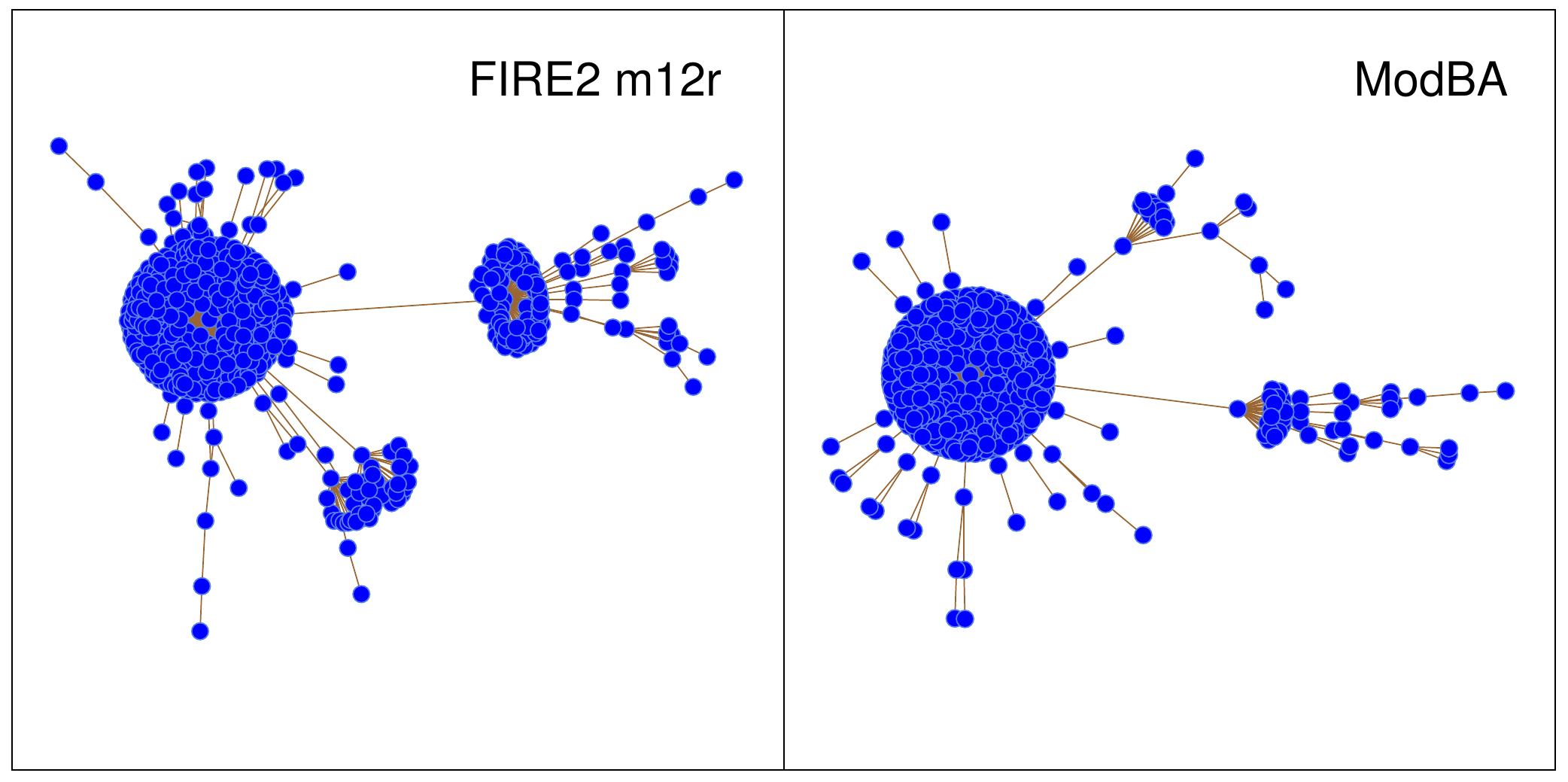}
  \caption{\label{fig:graph1} Visual comparison between a graph constructed from FIRE2 m12r (left) and an analogous one generated using the ModBA model with the kernel Eq.~\ref{eq:kernel2} (right). Both graphs have $N=600$ nodes.}
\end{figure}

Fig.~\ref{fig:graph1} shows visual comparison between a graph constructed from FIRE2 m12r and an analogous graph generated using the kernel~\ref{eq:kernel2}. Our ModBA model reproduces the structural properties of the simulated halo system, including the presence of a central hub and subclusters. The largest subcluster shown in the left panel represents a simulated analog of the Large Magellanic Cloud, the most massive satellite galaxy (subhalo) of the Milky Way~\cite{2020MNRAS.491.1471S}. See App.~\ref{app:free} for more examples and the criteria for choosing the analogous graph generated using the ModBA model. 

\begin{figure}[htbp]
  \centering
  \includegraphics[height=7.2cm]{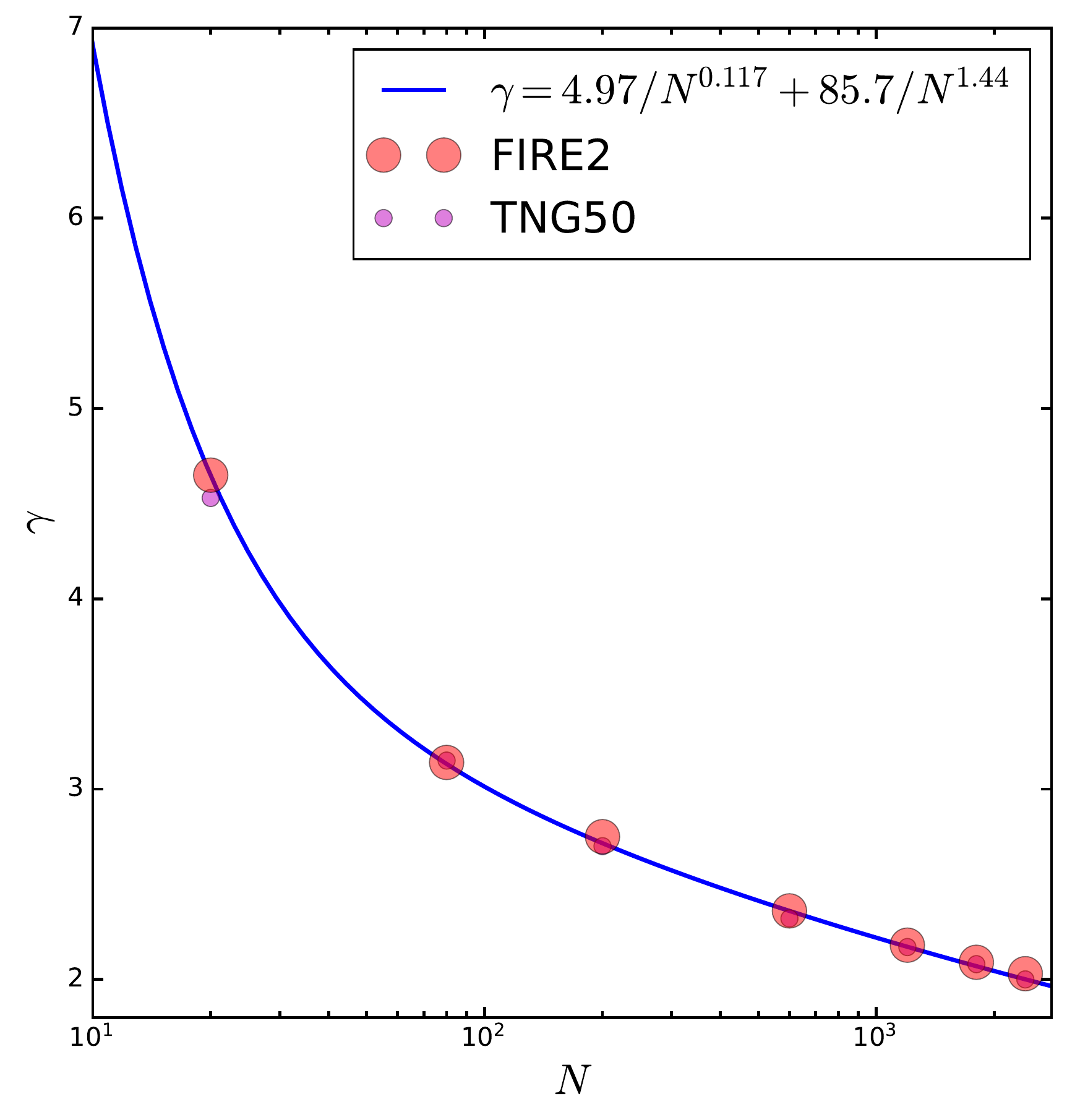}
  \caption{\label{fig:gamma} The $\gamma$ parameter as a function of the number of nodes $N$ in a graph. The calibrated values  using the FIRE2 (red) and TNG50 (magenta) simulations are shown on top of a fitted curve (blue).}
\end{figure}

{\noindent\bf Calibration of the kernel parameter $\gamma$.}
We provide details about the calibration of $\gamma$ as a function of $N$ in Eq.~\ref{eq:kernel2}. We first construct graphs for the halo systems from the FIRE2 and TNG50 simulations. For each constructed graph with a given number of total nodes $N$, we compute the number of higher-order subhalos $N_{\rm ho}$ as $N_{\rm ho} = N-N_{\rm hub}-1$, where ``1'' accounts for the main halo, and $N_{\rm hub}$ is the number of subhalos {\it directly} hosted by the main halo, i.e., the degree of the max degree node (central hub) in a graph. We then take the average over the number of reconstructed graphs with same $N$ and obtain $\left<N^{\rm FIRE2}_{\rm ho}\right>$ and $\left<N^{\rm TNG50}_{\rm ho}\right>$, where $\braket{...}$ refers to an average over the systems under consideration. As the next step, for given $N$, we generate a large number of graphs using our ModBA model with the kernel Eq.~\ref{eq:kernel2}, while varying the $\gamma$ value, until $\left<N^{\rm ModBA}_{\rm ho}\right>=\left<N^{\rm FIRE2}_{\rm ho}\right>$ or $\left<N^{\rm TNG50}_{\rm ho}\right>$. The calibration results are summarized in Tables~\ref{tab:tab1} and~\ref{tab:tab2}, as well as Fig.~\ref{fig:gamma}.

\begin{table}[bthp]
\begin{center}
\begin{tabular}{c|ccc|cc}
\hline
\hline
N   & $\left<N^{\rm FIRE2}_{\rm ho}\right>$ & $\# {\rm~of~graphs}$ & $\sigma^{\rm FIRE2}_{\rm ho}$ & $\sigma^{\rm ModBA}_{\rm ho}$ & $\gamma$ \\
\hline
$20  $& $1.55\pm 0.36$     & $11$     & $1.21$   &  $1.87$        &  $4.65$  \\
$80  $& $10.0\pm 2.4$      & $11$     & $8.02$   &  $8.75$        &  $3.14$  \\
$200 $& $26.1\pm 5.5$      & $11$     & $18.3$   &  $21.4$        &  $2.75$   \\
$600 $& $88.5\pm 16 $      & $11$     & $52.0$   &  $66.4$        &  $2.36$   \\
$1200$& $193 \pm 35 $      & $11$     & $116 $   &  $151 $        &  $2.18$  \\
$1800$& $301 \pm 53$       & $11$     & $177 $   &  $229 $        &  $2.09$   \\
$2400$& $409 \pm 72$       & $11$     & $239 $   &  $274 $        &  $2.03$  \\
\hline
\hline
\end{tabular}
\caption{\label{tab:tab1} The $\gamma$ values calibrated using the FIRE2 simulation.}
\end{center}
\end{table}

\begin{table}[bthp]
\begin{center}
\begin{tabular}{c|ccc|cc}
\hline
\hline
N   & $\braket{N^{\rm TNG50}_{\rm ho}}$ & $\# {\rm~of~graphs}$ & $\sigma^{\rm TNG50}_{{\rm ho}}$ & $\sigma^{\rm ModBA}_{N_{\rm ho}}$ & $\gamma$ \\
\hline
$20$  & $1.677\pm 0.042$     & $2004$ & $1.89$    & $1.98$&  $4.53 $  \\
$80$  & $9.67  \pm 0.35$      & $ 501$ & $7.77$    & $8.55$&  $3.15 $  \\
$200$ & $28.5\pm 1.2$        & $ 217$ & $18.0$    & $23.2$&  $2.70 $   \\
$600$ & $101.5 \pm 7.3$       & $  67$ & $60.0$    & $69.3$&  $2.32 $   \\
$1200$& $201 \pm 17$         & $  32$ & $94.4$    & $151$ &  $2.17 $  \\
$1800$& $323 \pm 29$         & $  26$ & $150 $    & $260$ &  $2.08 $   \\
$2400$& $499   \pm  51$        & $  18$ & $216 $    & $347$ &  $2.00 $  \\
\hline
\hline
\end{tabular}
\caption{\label{tab:tab2} The $\gamma$ values calibrated using the TNG50 simulation.
}
\end{center}
\end{table}
   
Take the $N=20$ case for example. For each of the $11$ simulated FIRE2 systems, we take $20$ most massive halos, including one main halo and $19$ subhalos of all orders that pass the mass cut, and find $\left<N^{\rm FIRE2}_{\rm ho}\right>=1.55$ and the standard deviation is $\sigma^{\rm FIRE2}_{\rm ho}=1.21$. The statistical uncertainty is estimated as $1.21/\sqrt{11}\approx0.36$. We find that for $\gamma=4.65$, our ModBA model can reproduce $\left<N^{\rm FIRE2}_{\rm ho}\right>=1.55$ with a standard deviation of $\sigma^{\rm ModBA}_{\rm ho}=1.87$. We repeat this procedure for other $N$ cases shown in Table~\ref{tab:tab1}. The results can be fitted with an empirical relation of $\gamma=4.97/N^{0.117}+85.7/N^{1.44}$, as shown in Fig.~\ref{fig:gamma}.

For the TNG50 simulation, there are $8800$ main halo systems that contain at least $1$ subhalo. Since the TNG50 sample is much larger than the FIRE2 one, we choose a slightly different way for calibrating $\gamma$. Instead of imposing a mass cut for choosing $N-1$ most massive subhalos, for each of the $8800$ systems, we reconstruct a graph for all of its associated halos, including all orders up to the resolution limit. Then we repeat the procedure discussed above to determine the $\gamma$ value, and the results are summarized in Table~\ref{tab:tab2}.
Consider the $N=20$ case, among the $8800$ systems, $2004$ of them have $N=20$ nodes, including $1$ main halo and $19$ subhalos of all orders up to the resolution limit. Accordingly, for $\gamma=4.53$, the graph model can produce the result from the simulation. We have also checked that the statistical uncertainties of $\gamma$ are less than $\sim 0.01$ for all cases calibrated with TNG50 because of the large number of graphs reconstructed from the TNG50 sample.

We see that the $\gamma$ values calibrated using the FIRE2 and TNG50 results agree remarkably well; see Fig.~\ref{fig:gamma}. This is mainly because $\left<N^{\rm FIRE2}_{\rm ho}\right>$ and $\left<N^{\rm TNG50}_{\rm ho}\right>$ are close to each other for all $N$ cases. In addition, since the attachment kernel Eq.~\ref{eq:kernel2} is
sensitive to $\gamma$ exponentially, a minor difference in $\gamma$ could accommodate the small difference in    
$\left<N_{\rm ho}\right>$ between two simulations.  Thus we conclude the empirical fitting relation $\gamma=4.97/N^{0.117}+85.7/N^{1.44}$ works for both simulations. Thus our ModBA model is robust. 

We have further checked the calibration of $\gamma$ using halos from the TNG-50-1 hydrodynamic simulation. For a fixed number of nodes $N$, the number of its associated graphs decreases, compared to TNG-50-1-Dark as listed in Table~\ref{tab:tab2}, as the feedback effects, such as tidal stripping from galaxies, can ``erase" subhalos. Nevertheless, the calibrated $\gamma$ values agree with those from the dark-matter-only simulation within $2\%$.

\begin{figure*}
  \centering
  \includegraphics[height=5.3cm]{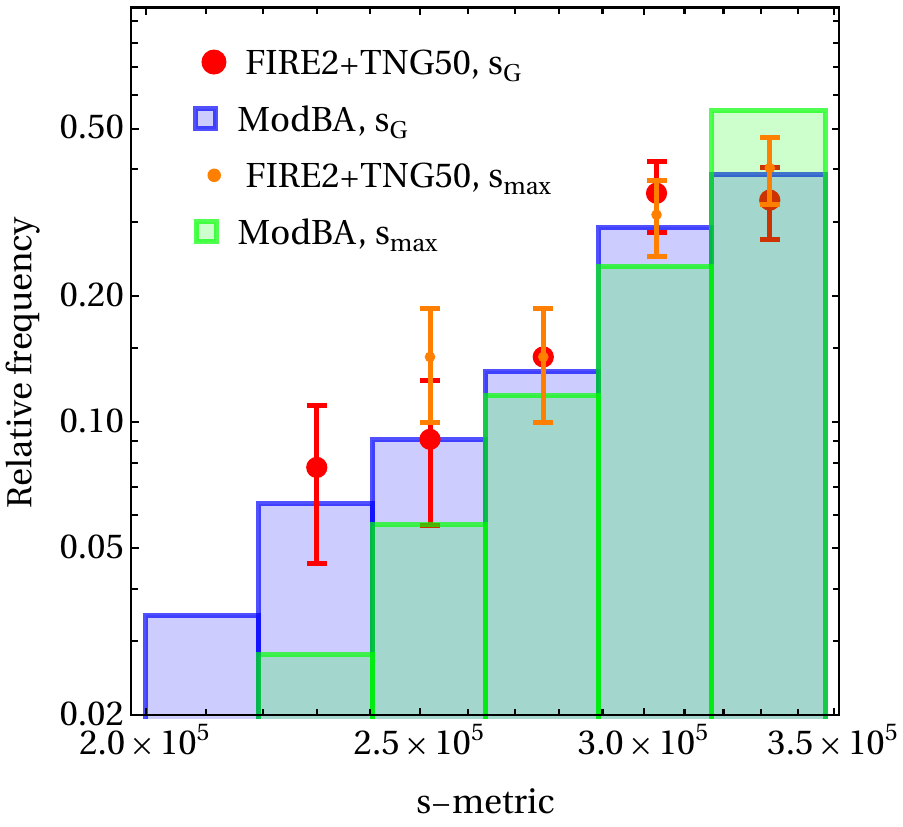}
  \includegraphics[height=5.2cm]{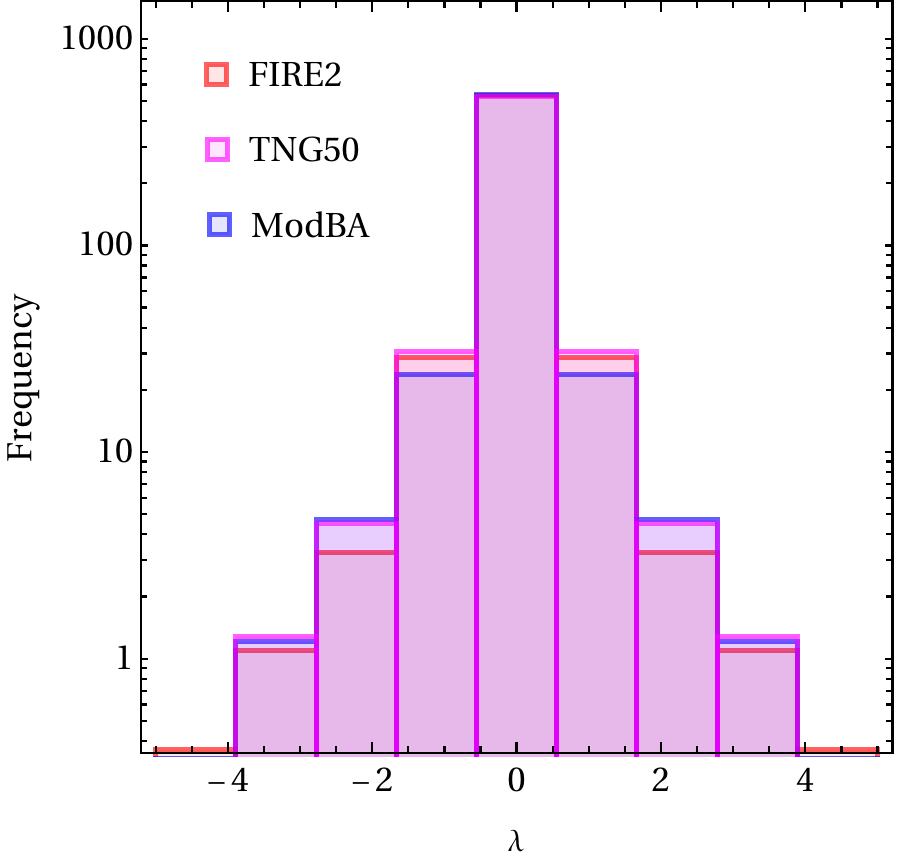}
  \includegraphics[height=5.2cm]{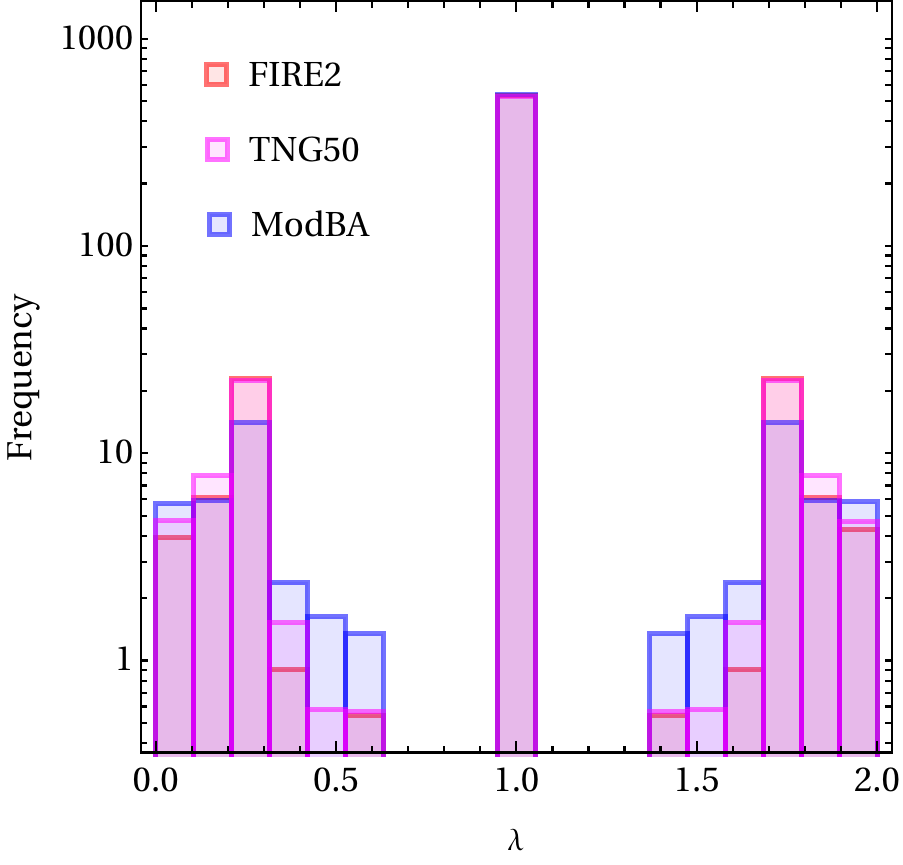}
  \caption{\label{fig:rst}
{\it Left}: The distribution of scale-free metric $s_{G}$ for the graphs from the FIRE2+TNG50 simulations (red dots with statistical errors), and those randomly generated using the ModBA model with the kernel Eq.~\ref{eq:kernel2} (blue rectangles). For comparison, the corresponding distribution of $s_{\max}$ is shown for the FIRE2+TNG50 (orange dots with errors) and ModBA (green rectangles) graphs. {\it Middle} and {\it Right:} the adjacency and Laplacian spectra, respectively, averaged over $11$ FIRE2 (red), $67$ TNG50 (magenta) and $100$ ModBA (blue) graphs with $N=600$ nodes. }
\end{figure*}

{\noindent\bf Scale-free and self-similar graphs.} Aside from the power-distribution shown in Fig.~\ref{fig:graph0}, Ref.~\cite{Li2005} proposes that one must check other important topological properties, such as self-similarity and universality, in evaluating a graph. Following~\cite{Li2005}, we use a structural metric to quantify the extent to which our constructed graphs are scale-free {\it and} self-similar. For a graph with edges forming a set $E$, the metric is defined as $s_G = \sum_{(i,j)\in E} n_i n_j$, where $(i,j)$ denotes an edge connecting nodes $i$ and $j$; see App.~\ref{app:free}. Fig.~\ref{fig:rst} (left) shows $s_G$ distributions for the graphs constructed from the FIRE2+TNG50 systems (red dots) and those generated using our ModBA model (blue rectangles). For both cases, the distribution peaks towards large values, indicating the graphs are scale-free. 

The value of $s_G$ can be maximized if one performs {\it degree-preserving} rewiring such that high-degree nodes are attached to other high-degree nodes~\cite{Li2005}. For a given degree sequence, $\{n_1,...,n_N\}$, we follow the algorithm in App. A of Ref.~\cite{Li2005} and reconstruct tree graphs that have the maximal $s_{G}$ value, $s_{\rm max}$.

\begin{figure*}[htbp]
  \centering
  \includegraphics[width=4.5cm]{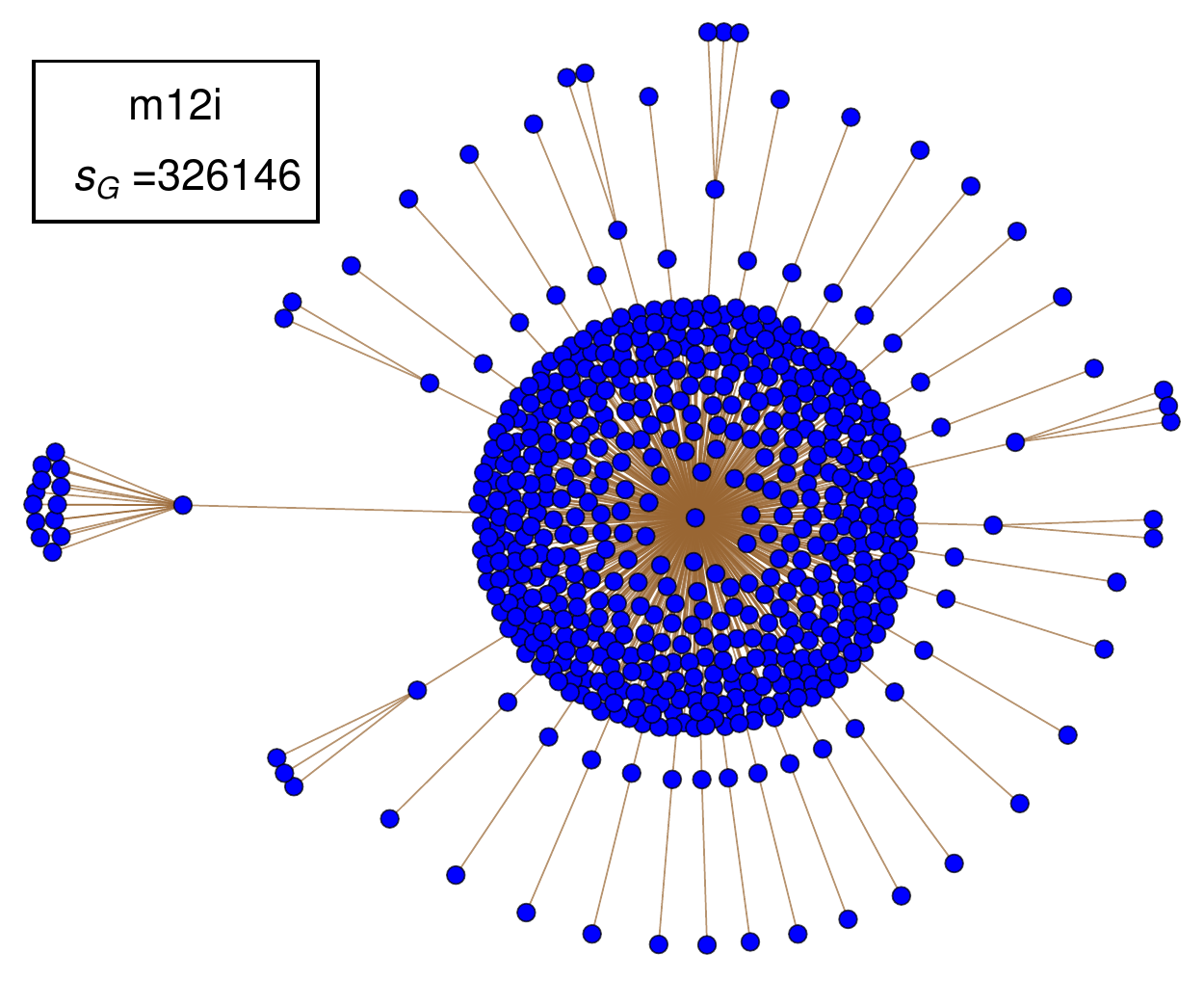}~~~~~~~~~
  \includegraphics[width=4.5cm]{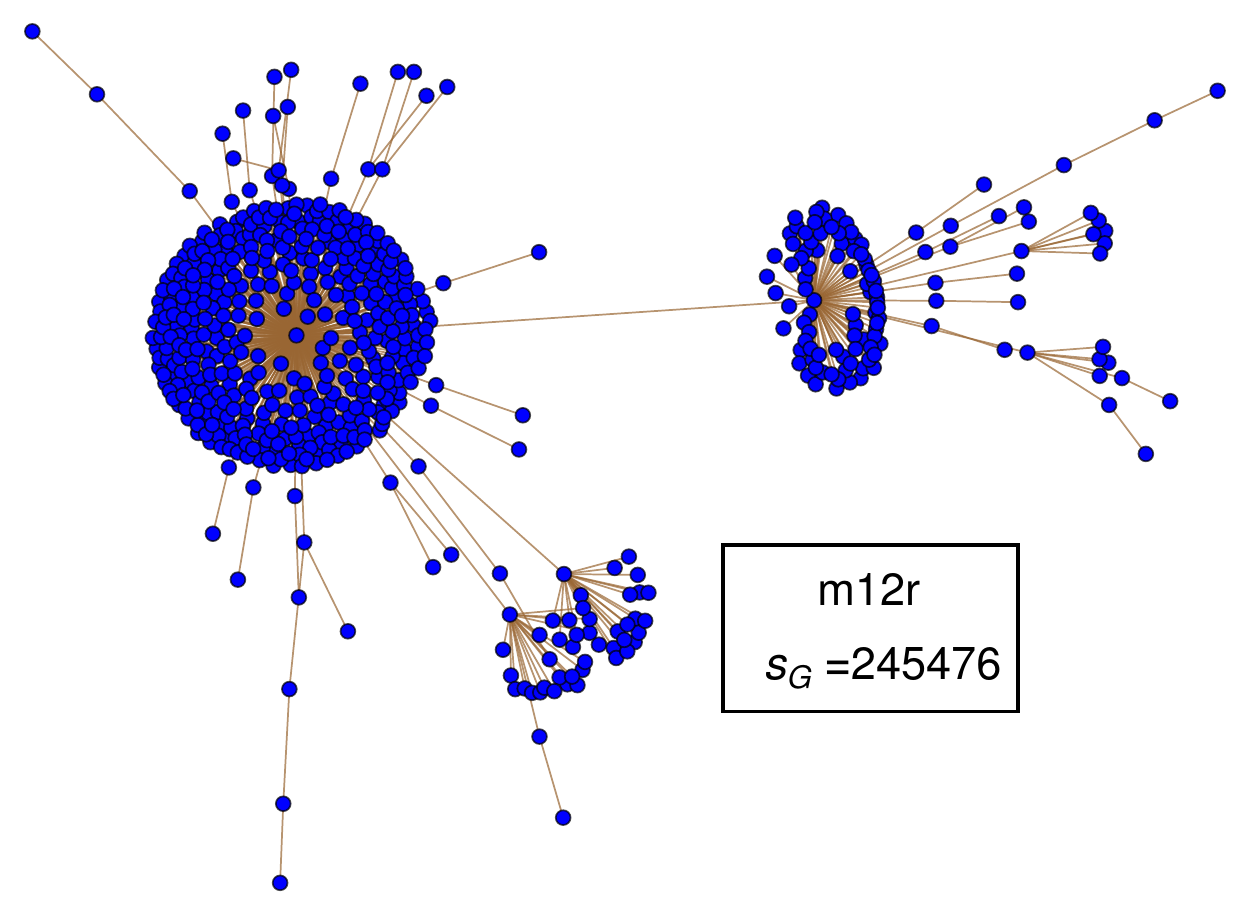}~~~~~~~~~
  \includegraphics[width=4.5cm]{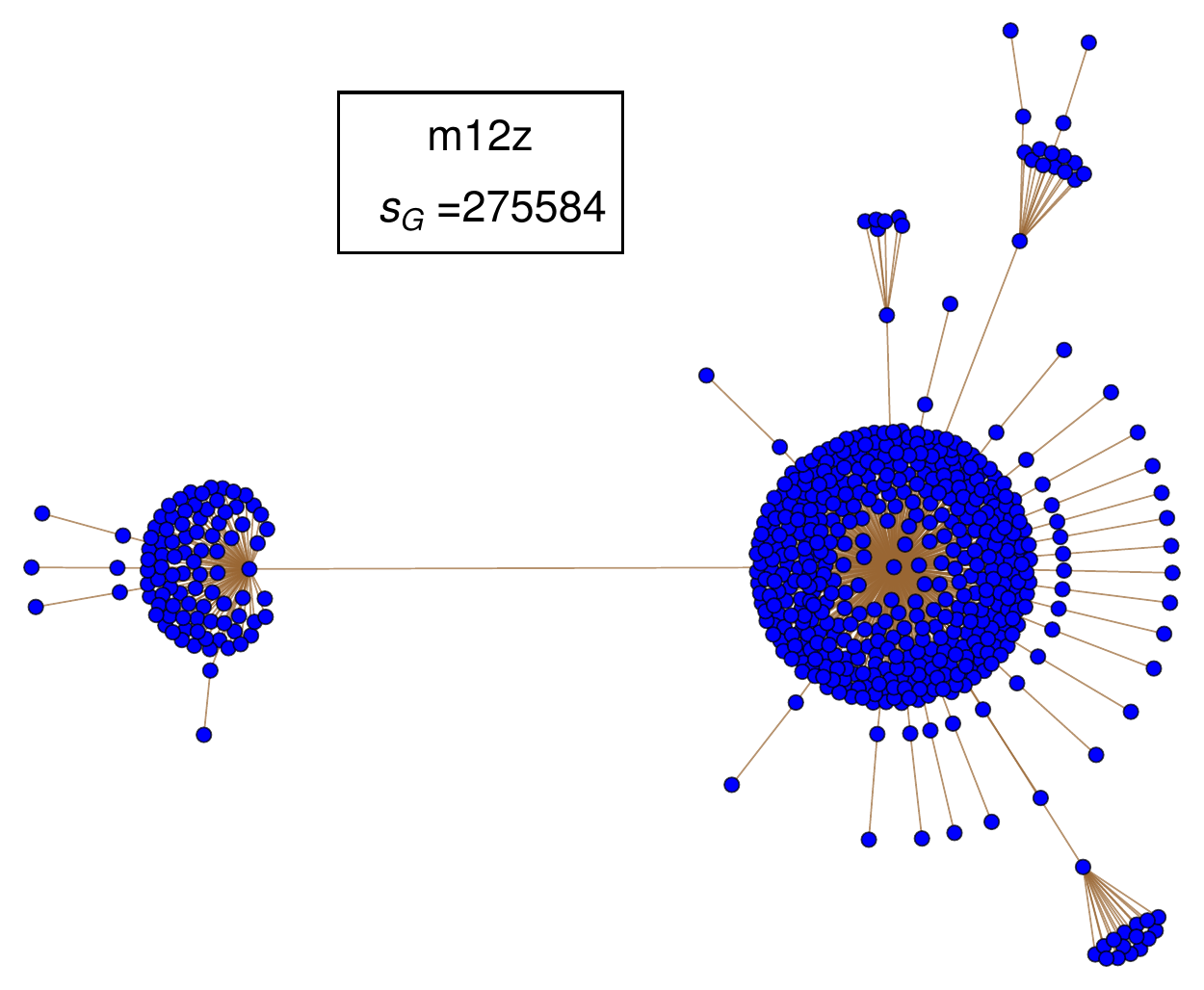} \\
  \includegraphics[width=4.5cm]{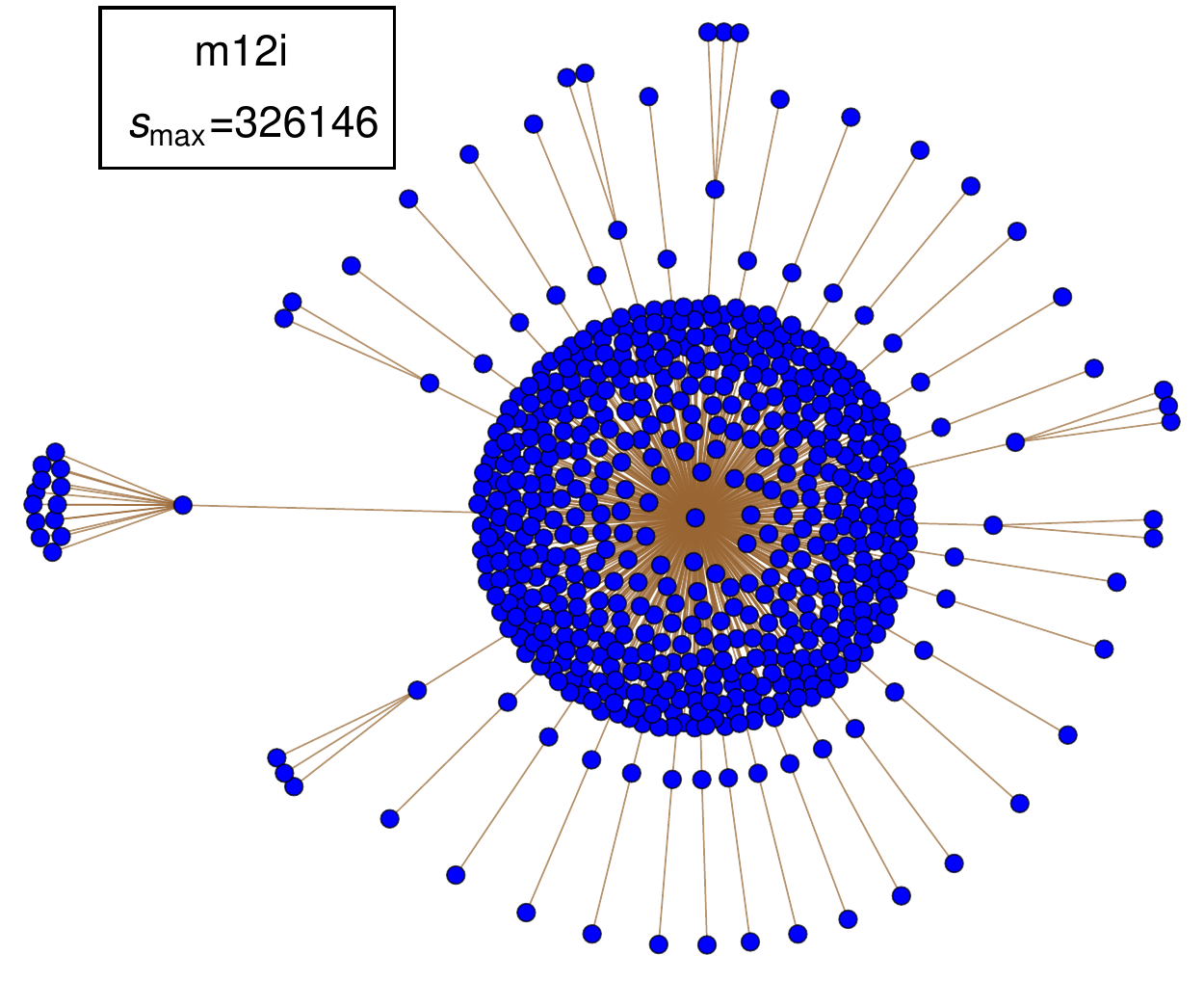}~~~~~~~~~
  \includegraphics[width=4.5cm]{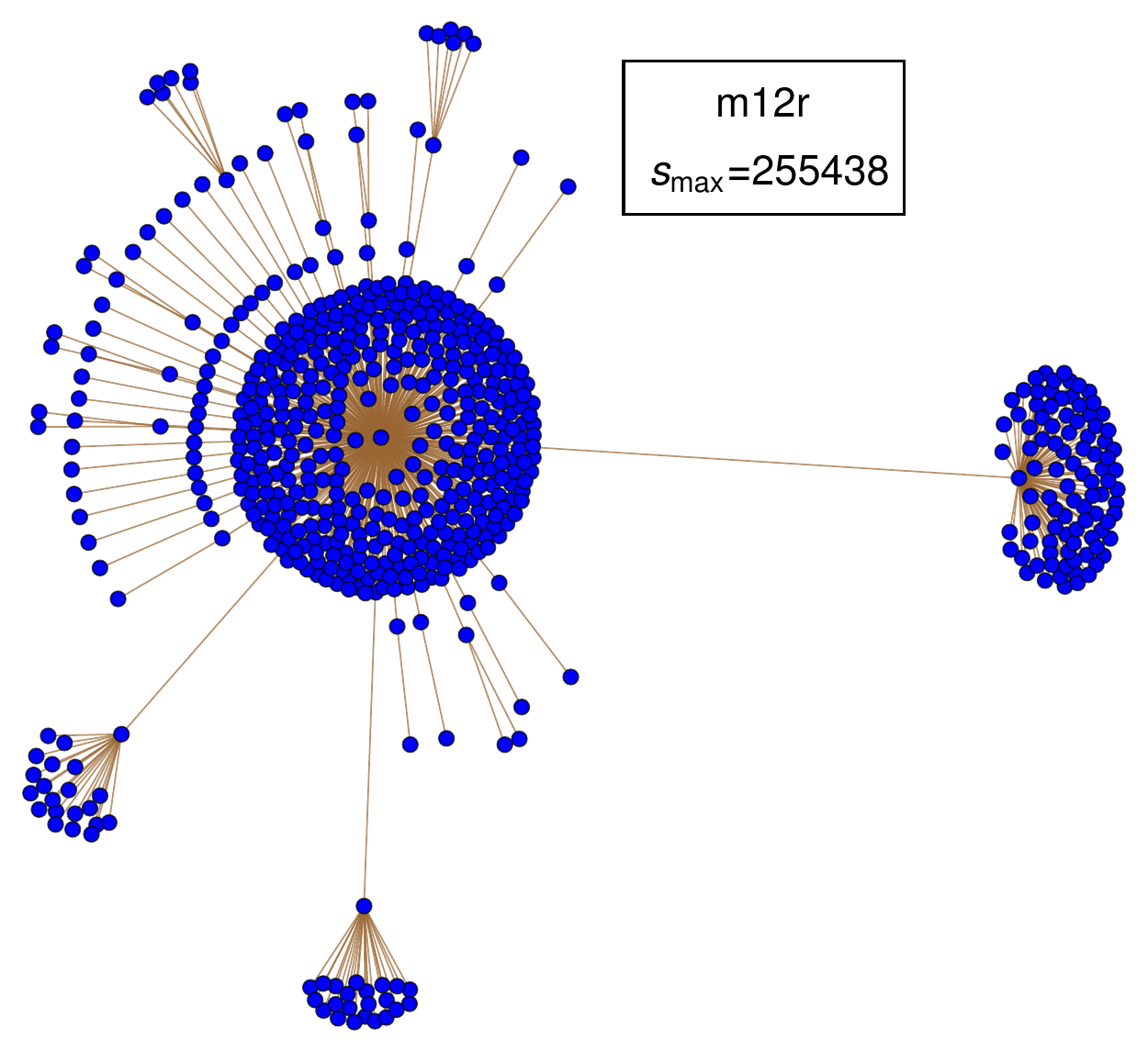}~~~~~~~~~
  \includegraphics[width=4.5cm]{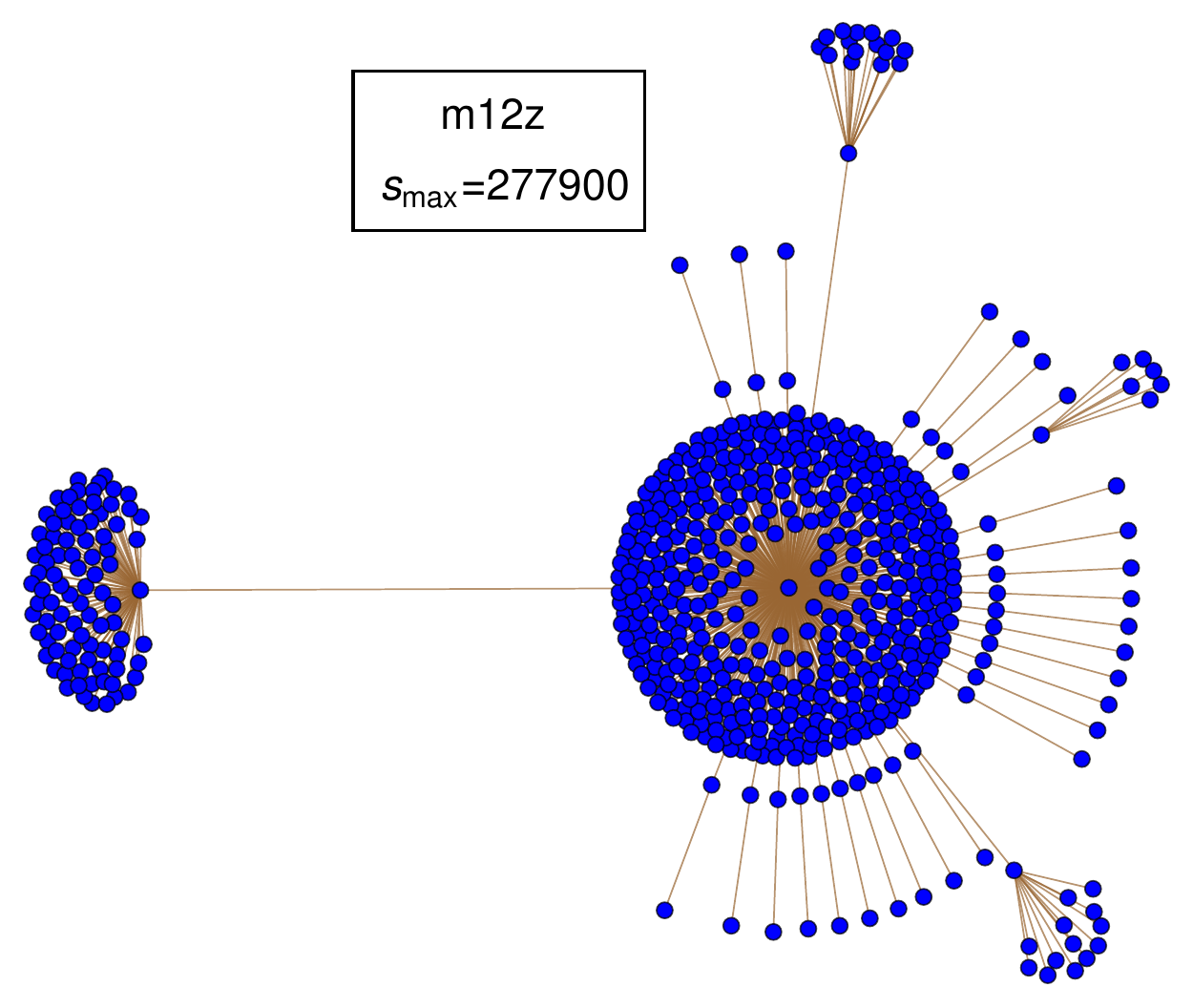} 
  \caption{\label{fig:self} {\it Top}: graphs constructed from three Milky Way-like systems in the FIRE2 project: m12i (left), m12r (middle), and m12z (right), together with their $s$-metric $s_G$ values. Each graph contains one main halo and $599$ most massive subhalos; the total number of nodes is $N=600$. {\it Bottom}: corresponding graphs of which $s_G$ reaches its maximum possible value $s_{\rm max}$, based on the degree-preserving rewiring algorithm in Ref.~\cite{Li2005}. Note the simulated m12i system happens to have $s_G=s_{\rm max}$, and hence the graphs on the top and bottom panels are identical.}
\end{figure*}

Fig.~\ref{fig:self} (top) shows graphs directly constructed from three simulated Milky Way-like systems in the FIRE2 project: m12i (left), m12r (middle), and m12z (right), as well as their graph metric $s_G$ values. For each graph, the total number of nodes is $N=600$, including one main halo and $599$ most massive subhalos. For each graph in the top panels, we use the degree-preserving rewiring algorithm in~\cite{Li2005} to find its corresponding graph configuration that has the maximum value of the metric, i.e., $s_{\rm max}$. If a graph (before rewiring) has $s_{\rm G}\approx s_{\rm max}$, it indicates that the graph is maximally scale-free and self-similar; see~\cite{Li2005} for details.

Fig.~\ref{fig:self} (bottom) shows corresponding graphs for m12i (left), m12r (middle), and m12z (right) after rewiring, and their $s_{\rm max}$ values. We see that for all three FIRE2 systems, the similarity between graph before and after rewiring is high, indicating that they are indeed maximally scale-free and self-similar. In particular for m12i, its graph before rewiring already has $s_G=s_{\rm max}$, and hence the configurations shown in the left top and bottom panels are identical.

Fig.~\ref{fig:rst} (left) shows $s_{\rm max}$ distribution for the rewired graphs from the FIRE2+TNG50 simulations (orange dots) and from the ModBA model (green rectangles). We see the $s_{\rm max}$ distribution is similar to the $s_G$ one. Quantitatively, the averaged ratios of $s_{G}/s_{\rm max}$ are $0.98$ and $0.93$ for the FIRE2+TNG50 and ModBA graphs, respectively. If a graph (before rewiring) has $s_{\rm G}\approx s_{\rm max}$, the graph is self-similar~\cite{Li2005}. Thus we have demonstrated that the cosmic structure formation is scale-free {\it and} self-similar according to the criteria of graph metrics, and our ModBA model can capture this important feature; see App.~\ref{app:free} for explicit examples. 

\begin{figure*}[htbp]
  \centering
  \includegraphics[width=6cm]{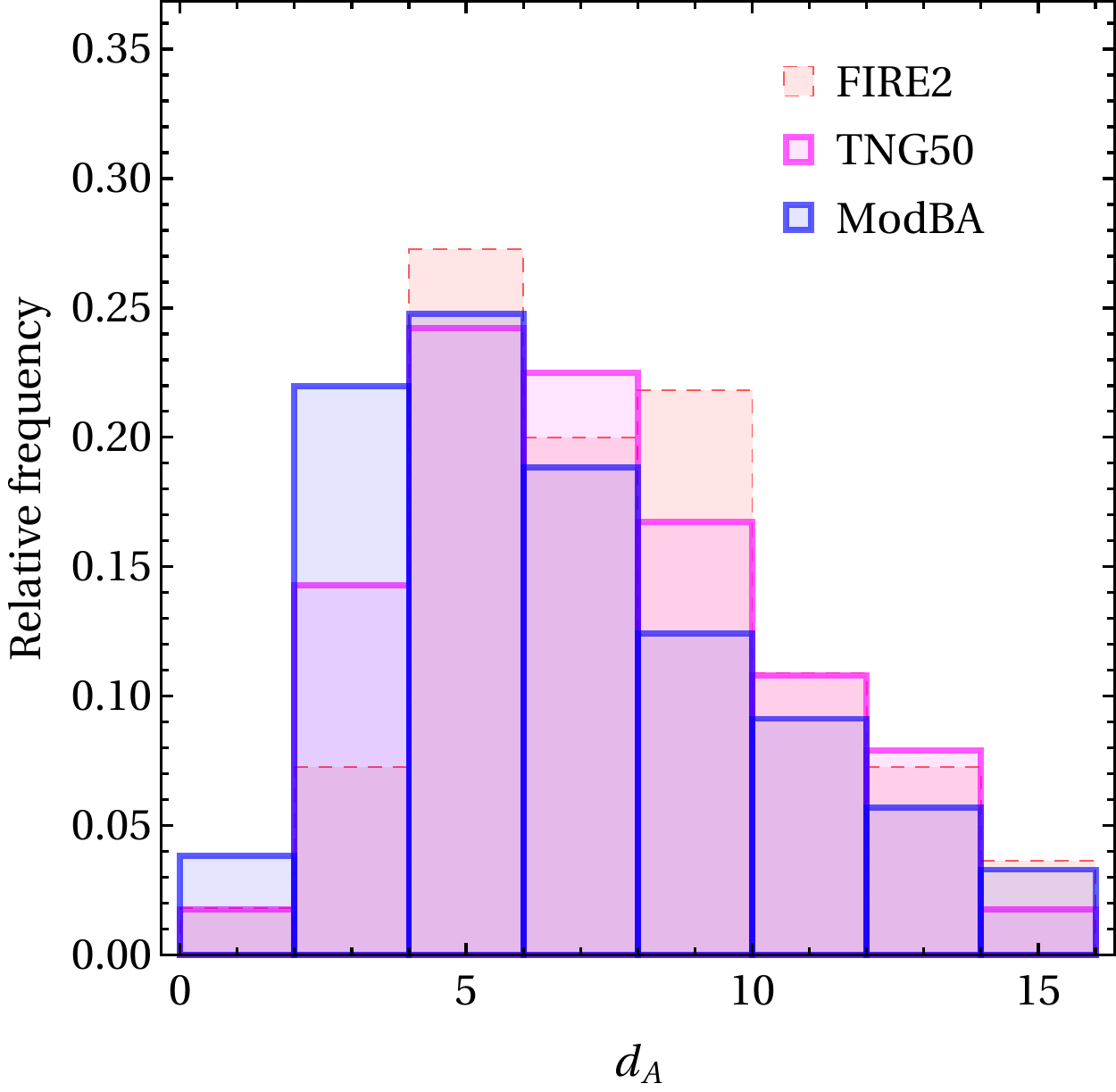}~~~~~
  \includegraphics[width=6cm]{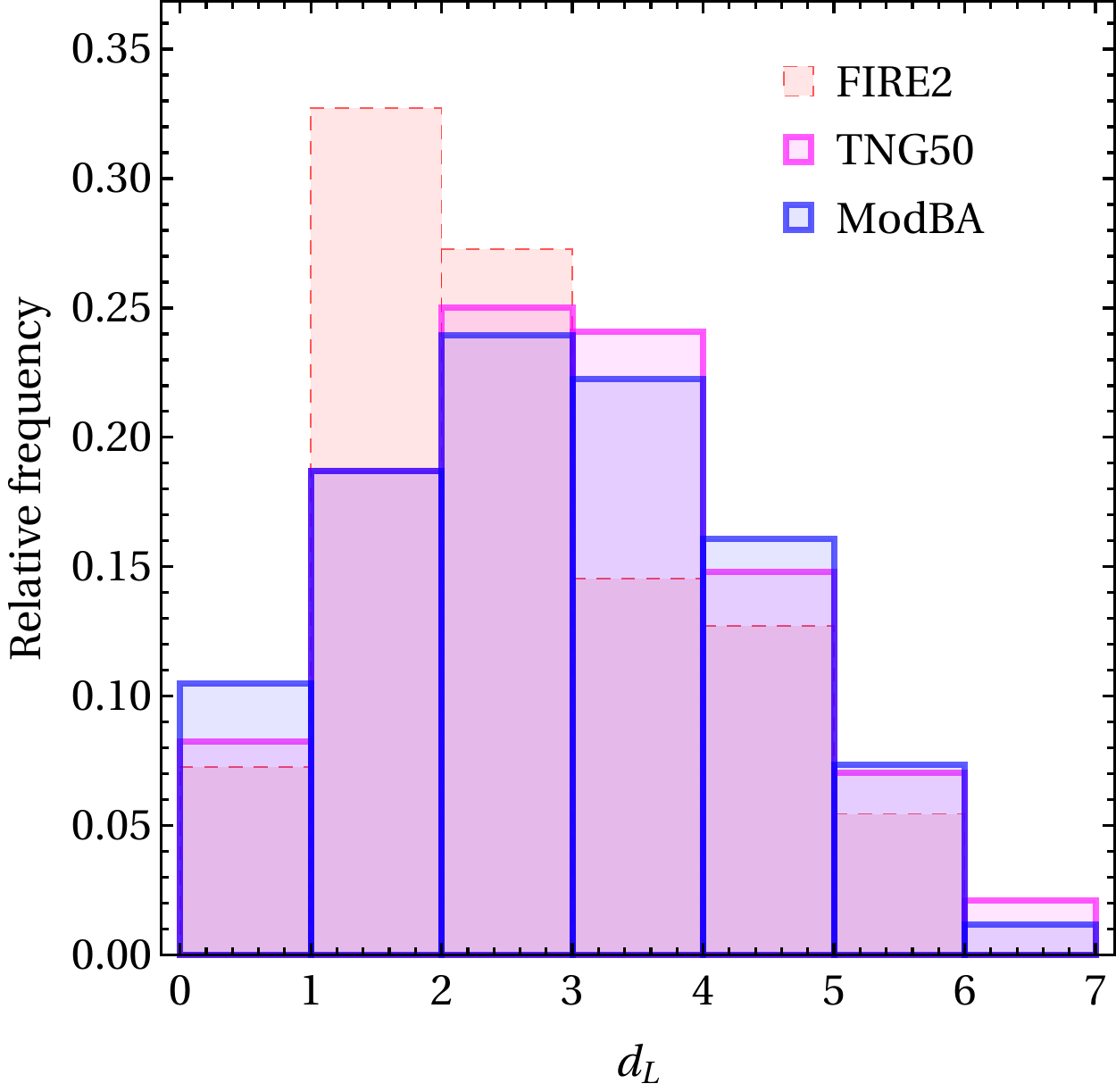}
  \caption{\label{fig:var}
  The distributions of graph distances $d_A$ (left) and $d_L$ (right) for the FIRE2, TNG50 and ModBA graphs. The FIRE2 distributions, denoted with the dashed line, are based on $11$ graphs, and they may suffer from relatively large statistical uncertainties.
}
\end{figure*}

{\noindent\bf Graph spectra.} 
We use the adjacency and normalized Laplacian matrices to describe the structure of a graph~\cite{Albert2002,Boccaletti2006}, where directions of graph edges are neglected.
For $N$ halos, including subhalos and the host, we take the $i-j$ and $j-i$ matrix element to $1$ when linking the $i^{th}$ and the $j^{th}$ halos. All the rest matrix elements are taken to be zero.
Based on the adjacency matrix, the normalized Laplacian matrix can be obtained as
\begin{eqnarray}
L =  \mathbb{1} - D^{-1/2} A D^{-1/2}, 
\end{eqnarray}
where $D$ is a diagonal degree matrix constructed by summing over elements of the adjacency matrix in each column and putting them to the corresponding diagonal positions.

Fig.~\ref{fig:rst} shows the spectra of the adjacency (middle) and normalized Laplacian (right) matrices of the FIRE2 (red), TNG50 (magenta) and ModAB (blue) graphs. The adjacency spectra are symmetric around zero and strongly localized around the hub. The pair of the largest eigenvalues are close to $\sqrt{N}$, which are not shown for clarity~\cite{PhysRevE.64.051903}. The eigenvalue of the normalized Laplacian matrix is in a range of $0\textup{--}2$. The corresponding spectra have a strong correlation to the topological features of a network~\cite{Banerjee2008,BANERJEE20092425,DeLange2014,868688,doi:10.1063/1.2008598,2010PhR...486...75F,2010JSMTE..10..020S,2009arXiv0910.3118B}: small and large eigenvalues are associated with the clustering and “bipartiteness” of graph substructures, respectively. The peak around one indicates that our graphs have a large central hub. Our ModAB model produces very similar spectra to those from the cosmological simulations. 
 
The comparisons in Fig.~\ref{fig:rst} are based on the overall feature of the spectra. We can further characterize those graphs using the graph distance.
Consider two connected tree graphs $G_1$ and $G_2$, and each contains $N$ nodes. We denote their eigenvalues of the adjacency matrix as $\lambda_{1,i}$ and $\lambda_{2,i}$, respectively, and the normalized distance between two graphs is computed as
\begin{eqnarray}
d_A(G_1,G_2) = \sqrt{\sum_{i=1}^{N}(\lambda_{1,i} - \lambda_{2,i})^2}.
\end{eqnarray}
Similarly, for the normalized Laplacian matrix, the graph distance is
\begin{eqnarray}
d_L(G_1,G_2) = \sqrt{\sum_{i=1}^{N}(\mu_{1,i} - \mu_{2,i})^2},
\end{eqnarray}
where $\mu_{1,i}$ and $\mu_{2,i}$ are eigenvalues of graphs $G_1$ and $G_2$, respectively.

Fig.~\ref{fig:var} shows the $d_A$ (left) and $d_L$ (right) distributions of the FIRE2 (red), TNG50 (magenta) and ModBA (blue) graphs, where the number of notes is $N=600$. We compute the graph distances among all distinct pairs within each case. There are $11$, $67$ and $100$ graphs for FIRE2, TNG50 and ModBA, respectively. We see the overall agreement is good among the three cases, although the FIRE2 distributions may suffer from relatively larger statistical uncertainties. In Table~\ref{tab:tab3}, we show the mean distances and their standard deviations for the three cases and again see good agreement.

\begin{table}[bthp]
\begin{center}
\begin{tabular}{c|cc|cc}
\hline
\hline
 &   $\braket{d_A}$       & $\sigma(d_A)$      &  $\braket{d_L}$  & $\sigma(d_L)$ \\
\hline
FIRE2          &    7.7        & 3.2                &   2.6   &  1.3   \\
TNG50          &   7.6         & 3.8                &   3.0   &  1.4   \\
ModBA          &   7.4         &     4.5            &   2.9   &   1.5  \\
\hline
\hline
\end{tabular}
\caption{\label{tab:tab3} The mean graph distances $\left<d_A\right>$, $\left <d_L\right>$, and their associated standard deviations $\sigma(d_A)$ and $\sigma(d_L)$ for FIRE2, TNG50, and ModBA graphs. }
\end{center}
\end{table}

{\noindent\bf Conclusions.} We have proposed a graph model to study the clustering of dark matter halos. In particular, we focused on subhalos from cosmological simulations of structure formation and constructed tree graphs that characterize the relationship between a subhalo and its host halo. Our model is based on preferential attachment and it successfully reproduce the clustering properties of the simulated halo systems. We also quantitatively demonstrated that cosmic structure formation is scale-free and self-similar. There are several related topics worthy of further investigations. We could extend our study to dark matter scenarios beyond the standard one, such as those with strong dark matter self-interactions~\cite{Tulin:2017ara,Adhikari:2022sbh} or damped matter power spectra~\cite{Cyr-Racine:2015ihg}, where the abundance of subhalos can be suppressed. In addition, other halo properties, such as shape and orientation, can be incorporated as weights in nodes and edges in our model, offering a natural framework for neural network studies, see recent examples~\cite{4700287,2016arXiv160902907K,2017arXiv171010903V,Zhou2020,Xu:2018xnz,Villanueva-Domingo:2021dun,Villanueva-Domingo:2021tit,Villanueva-Domingo:2022rvn,Jagvaral2022,Makinen:2022jsc}.

\section*{Acknowledgments}
We thank Ethan Nadler and Philip Tanedo for useful discussions. We acknowledge the FIRE and IllustrisTNG collaborations for making their simulation data publicly available. This work was supported by the John Templeton Foundation under Grant ID \#61884, the U.S. Department of Energy under Grant No. de-sc0008541, and NASA under Grant No. 80NSSC20K0566. The opinions expressed in this publication are those of the authors and do not necessarily reflect the views of the John Templeton Foundation. 

\bibliography{reference}

\appendix

\section{Comparison between the two kernels}
\label{app:comparison}

\begin{figure*}[htbp]
  \centering
  \includegraphics[height=5cm]{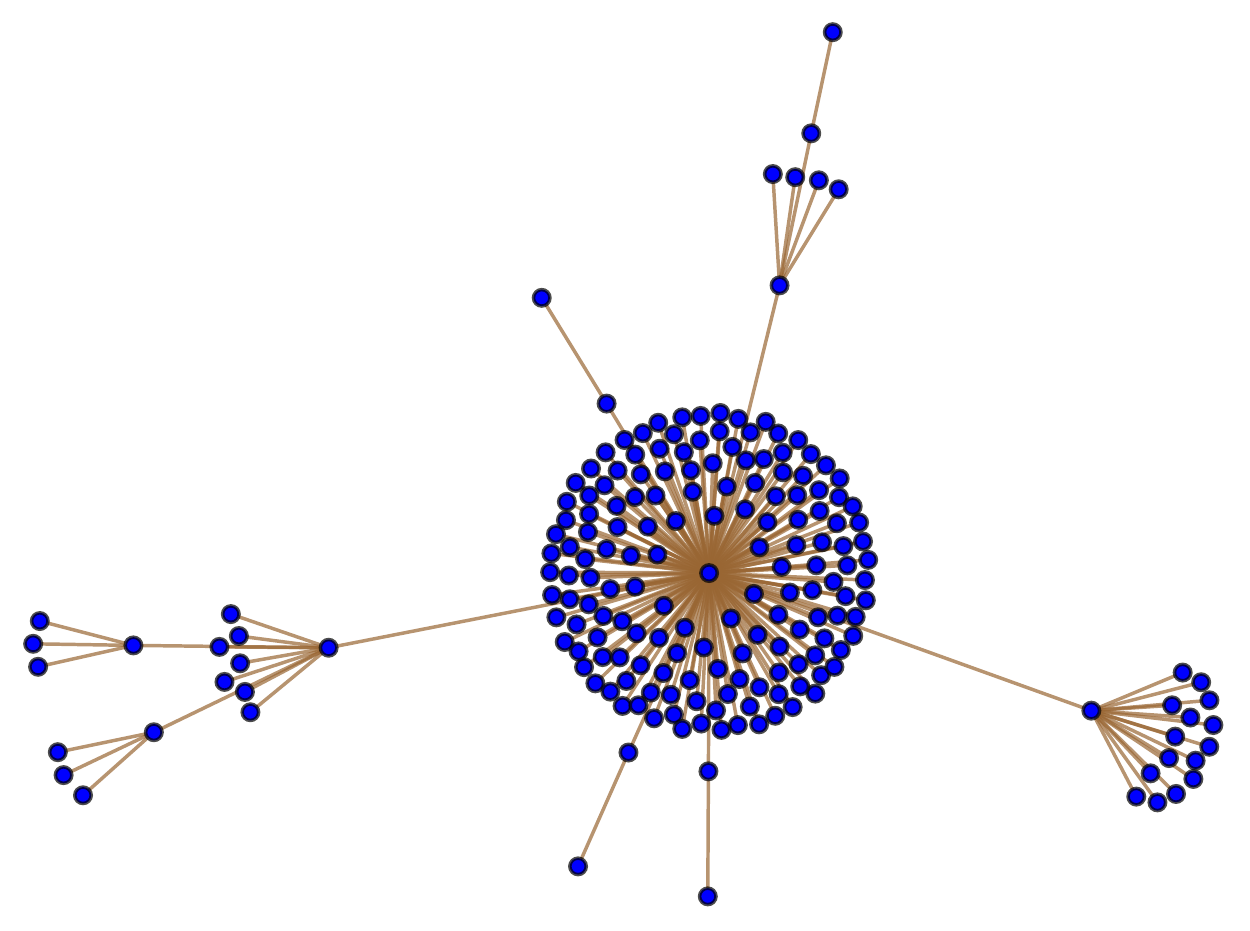}~~~~~~~~~~~~~~~~~~~~~
 \includegraphics[height=5cm]{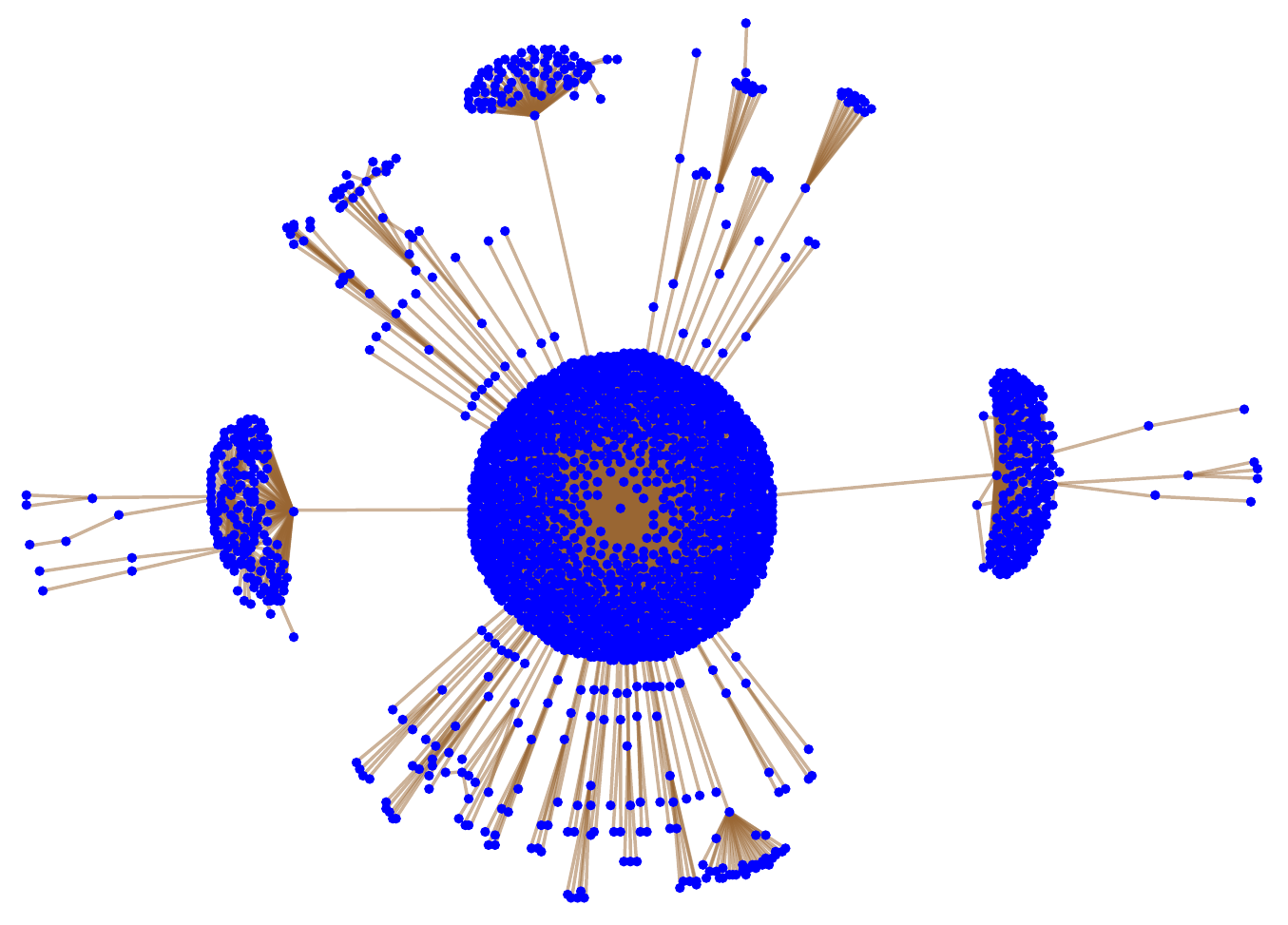}\\
  \includegraphics[height=4cm]{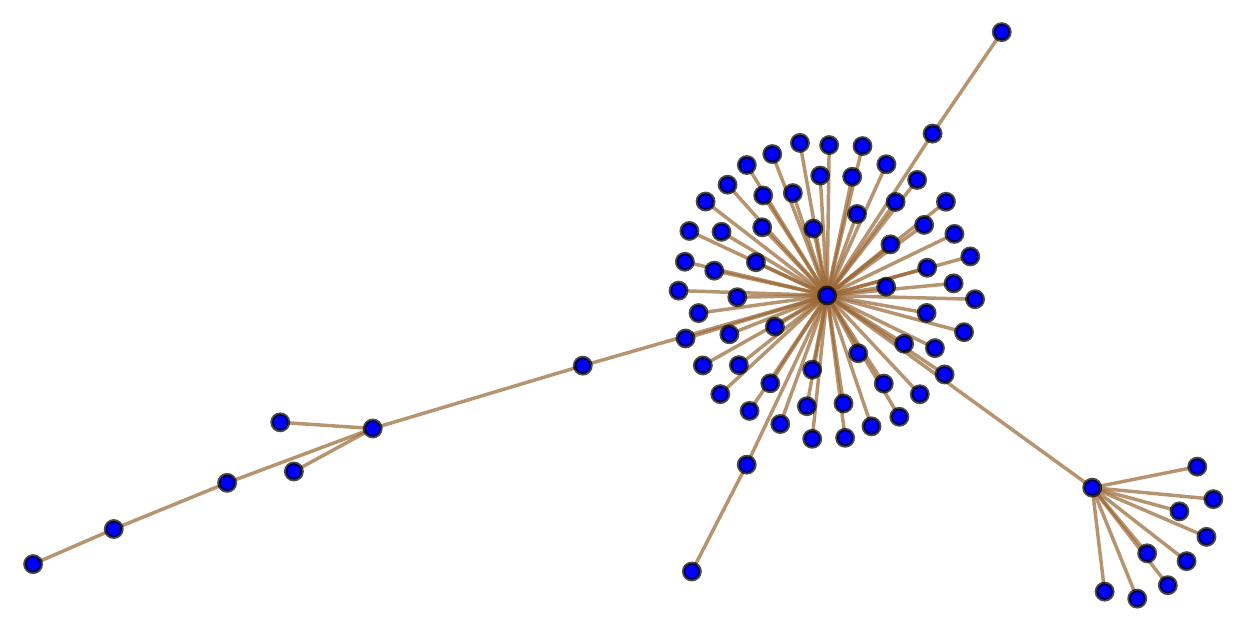}~~~~~~~~~~~~~~~~~~~~~
 \includegraphics[height=6cm]{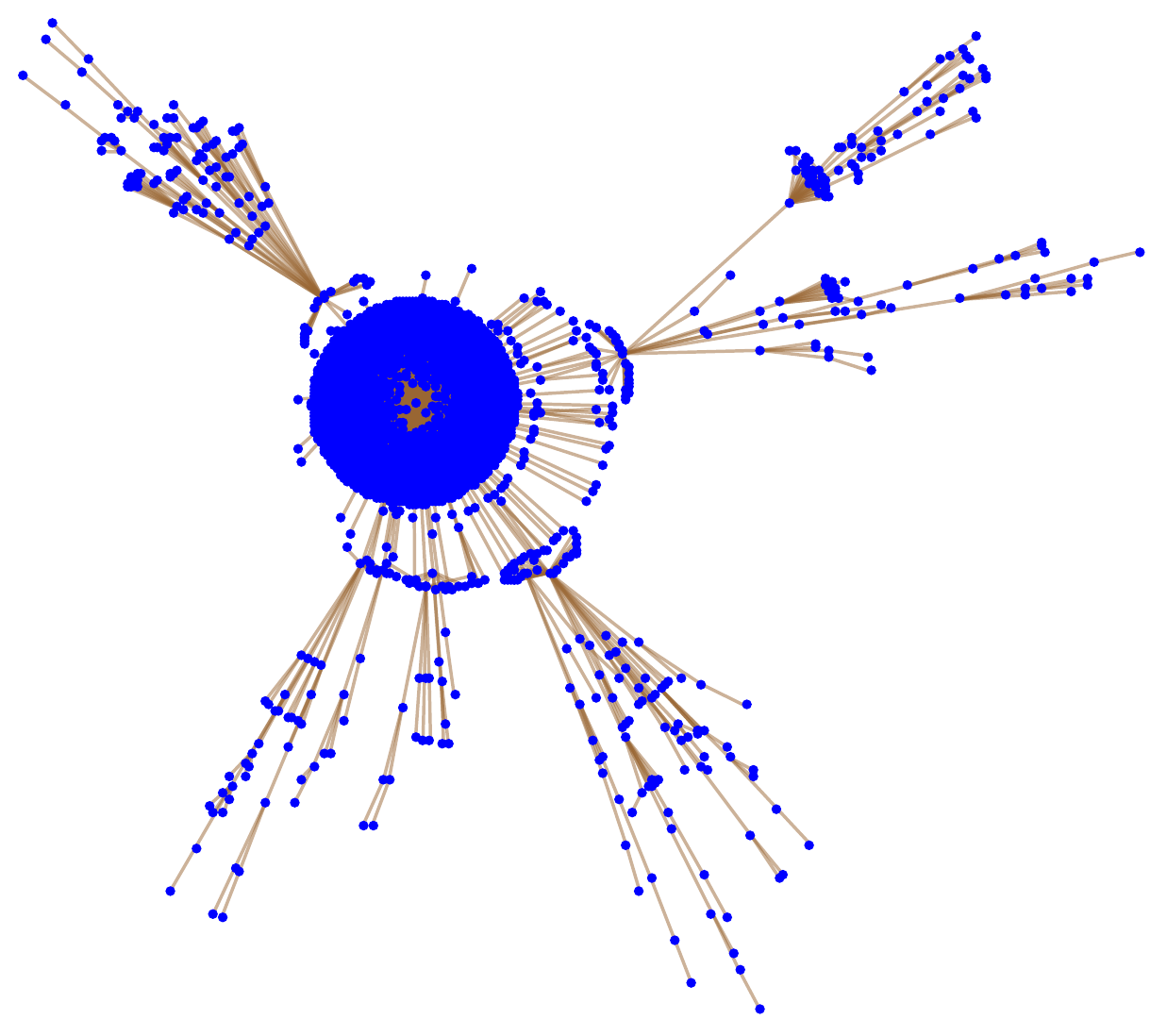}
  \caption{\label{fig:k1k2} Example graphs generated using the attachment kernels Eq.~\ref{eq:kernel1} (top) and Eq.~\ref{eq:kernel2} (bottom) with $N=80$ (left) and $N=2400$ (right).}
\end{figure*}

It is insightful to explicitly check whether the kernel Eq.~\ref{eq:kernel1} can produce graphs to model individual simulated halo systems for a given number of nodes $N$. Fig.~\ref{fig:k1k2} shows graphs randomly generated using the attachment kernels Eq.~\ref{eq:kernel1} (top) and Eq.~\ref{eq:kernel2} (bottom) with $N=80$ (left) and $N=2400$ (right). We see that for fixed $N$ the graphs using the two kernels look similar overall. However, there is a subtle difference, i.e., the graphs using the kernel Eq.~\ref{eq:kernel1} contain more nodes whose in-degree is high, compared to those using the kernel Eq.~\ref{eq:kernel2}. The difference becomes more significant as $N$ increases. For $N=2400$, the graph in the top panel contains three apparent ``subclusters'' that have significant high numbers of nodes, aside from the central hub, while the corresponding graph in the bottom panel does not have such sizable subclusters. 

Quantitatively, for $N=2400$, the graph using the kernel Eq.~\ref{eq:kernel1} has $N_{\rm hub}=1709$, and hence $N_{\rm ho}=N-N_{\rm hub}-1=690$. The graph using the kernel Eq.~\ref{eq:kernel2} has $N_{\rm hub}=1910$, and $N_{\rm ho}=489$. From Tables~\ref{tab:tab1} and \ref{tab:tab2}, we see that the averaged $N_{\rm ho}$ values are $409$ and $499$ for the graphs constructed from the FIRE2 and TNG50 simulations, respectively. Thus  the kernel Eq.~\ref{eq:kernel1} predicts smaller $N_{\rm hub}$, while larger $N_{\rm ho}$, compared to the simulations. We have tested this with a large sample of graphs and different numbers of nodes and found the tension holds, although for small $N$, statistical fluctuations could be large for individual graphs randomly generated. In addition, we have varied other parameters in the kernel Eq.~\ref{eq:kernel1} and found that it is not possible to achieve both power law $N^{-2}_{\rm sub}$ and $N_{\rm ho}$ found in the N-body simulations simultaneously. 

To fix the mismatch, we introduce the kernel Eq.~\ref{eq:kernel2}, which is linear for small  $n_j$, while having an enhancement as $n^{\gamma}_j$ for larger $n_j$. This enhancement factor helps us reduce $N_{\rm ho}$ and increase $N_{\rm hub}$, relative to the linear attachment. This is because the factor preferentially increases the probability of attaching a node to the host halo (i.e., hub), not to a “subscluster,” as the former has the highest numbers of subhalos directly attached. Physically, one could imagine that the halos inside the ``subclusters" are released to the main halo due to tidal stripping (as their direct host halo being stripped), and they become subhalos of the main halo, leading to an increase of $N_{\rm hub}$ and a decrease of $N_{\rm ho}$.

\section{Example graphs demonstrating scale-free and self-similar features}
\label{app:free}

In this section, we show example graphs to further demonstrate that our ModBA model can capture the scale-free and self-similar features of the cosmic structure formation.

\begin{figure*}[htbp]
  \centering
  \includegraphics[width=4.5cm]{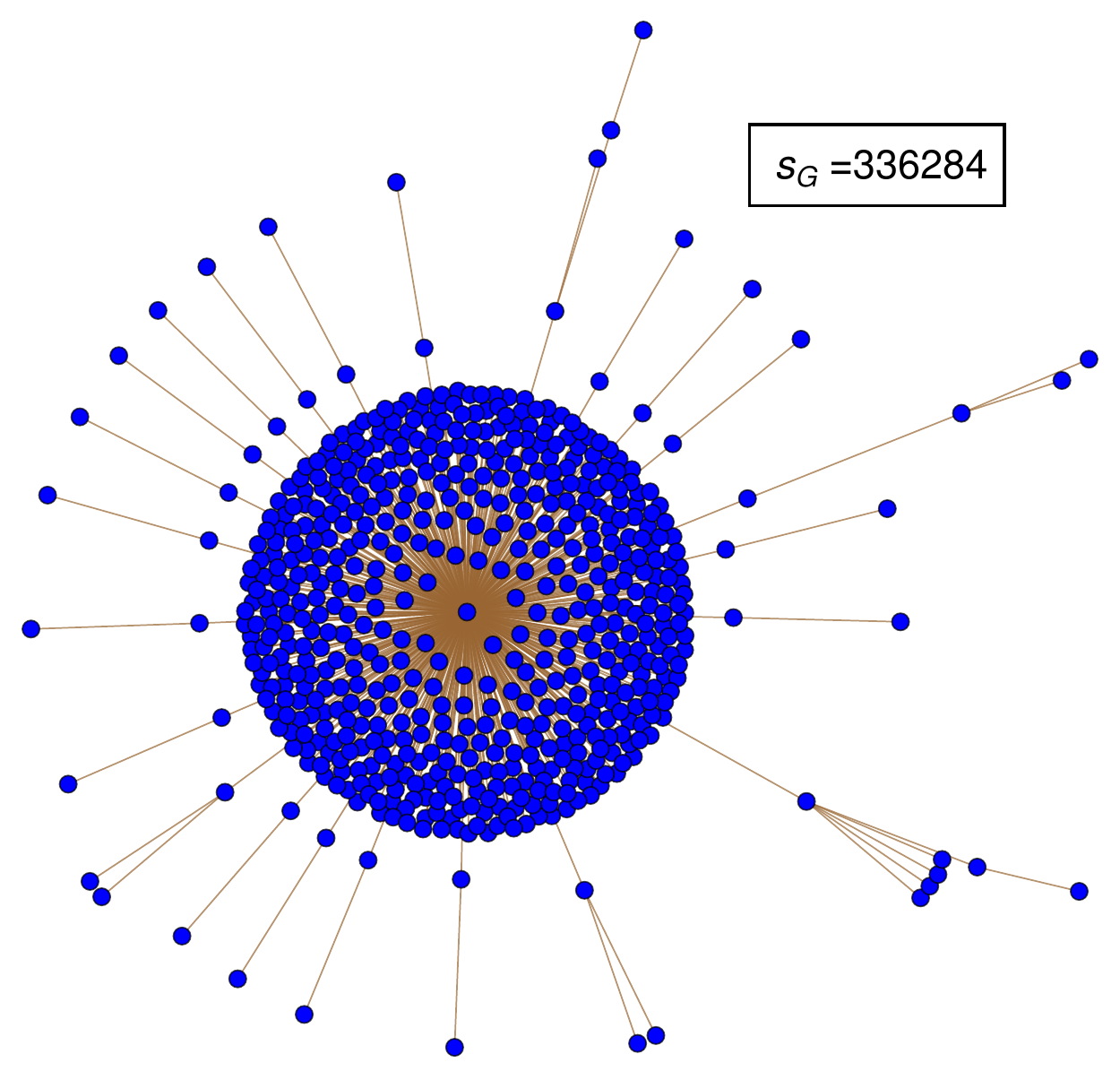}~~~~~~~~~
  \includegraphics[width=4.5cm]{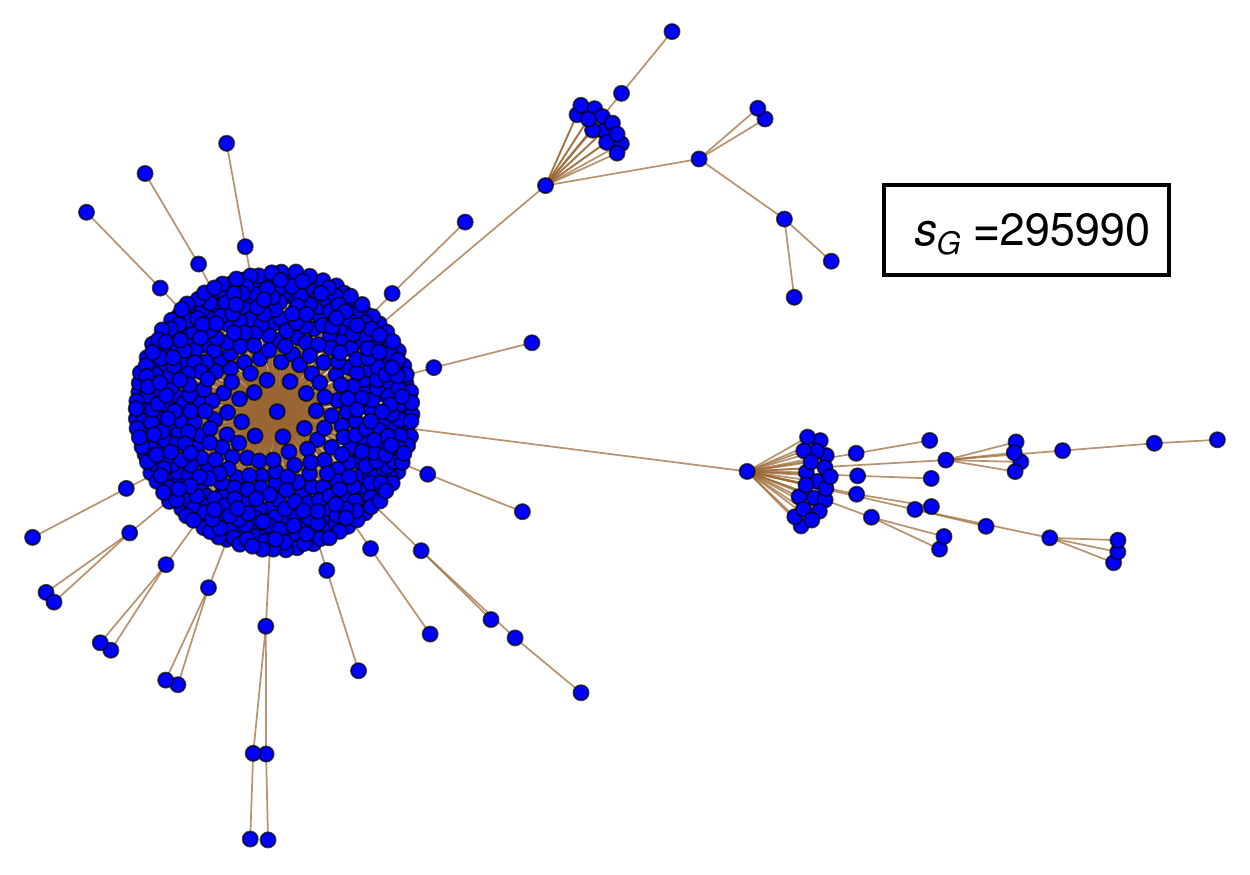}~~~~~~~~~
  \includegraphics[width=4.5cm]{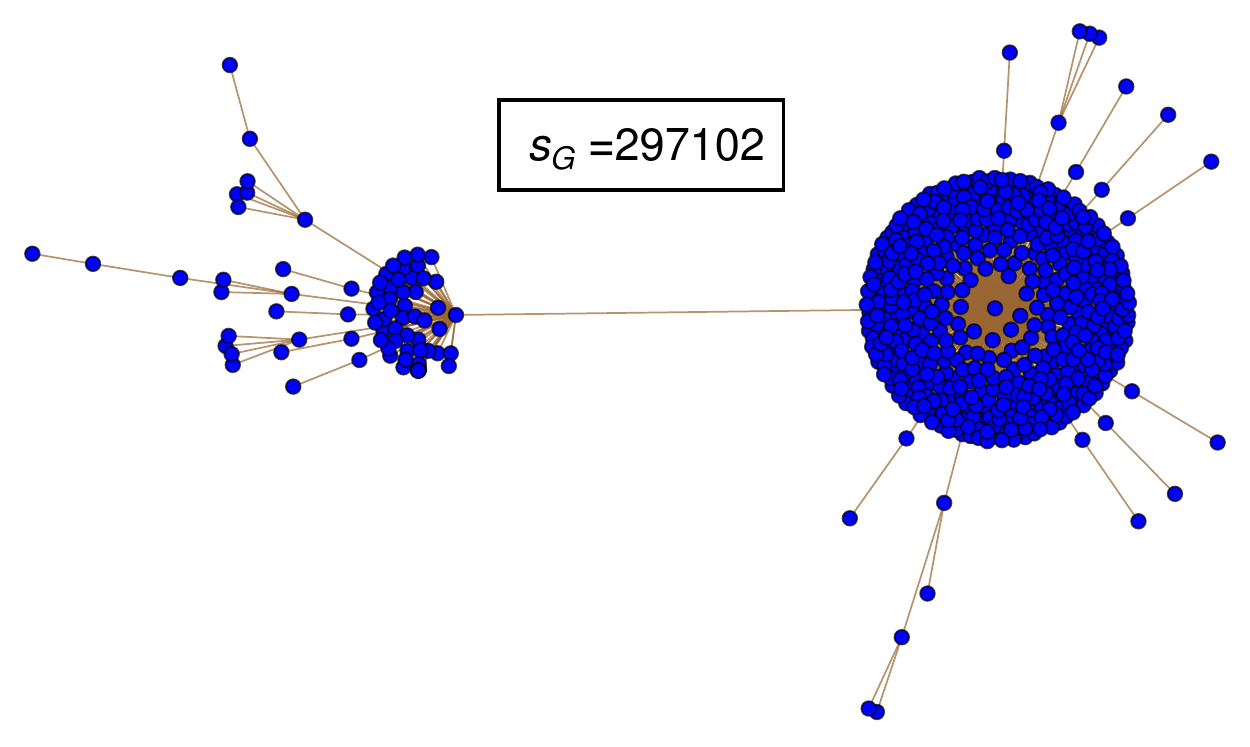}\\
  \includegraphics[width=4.5cm]{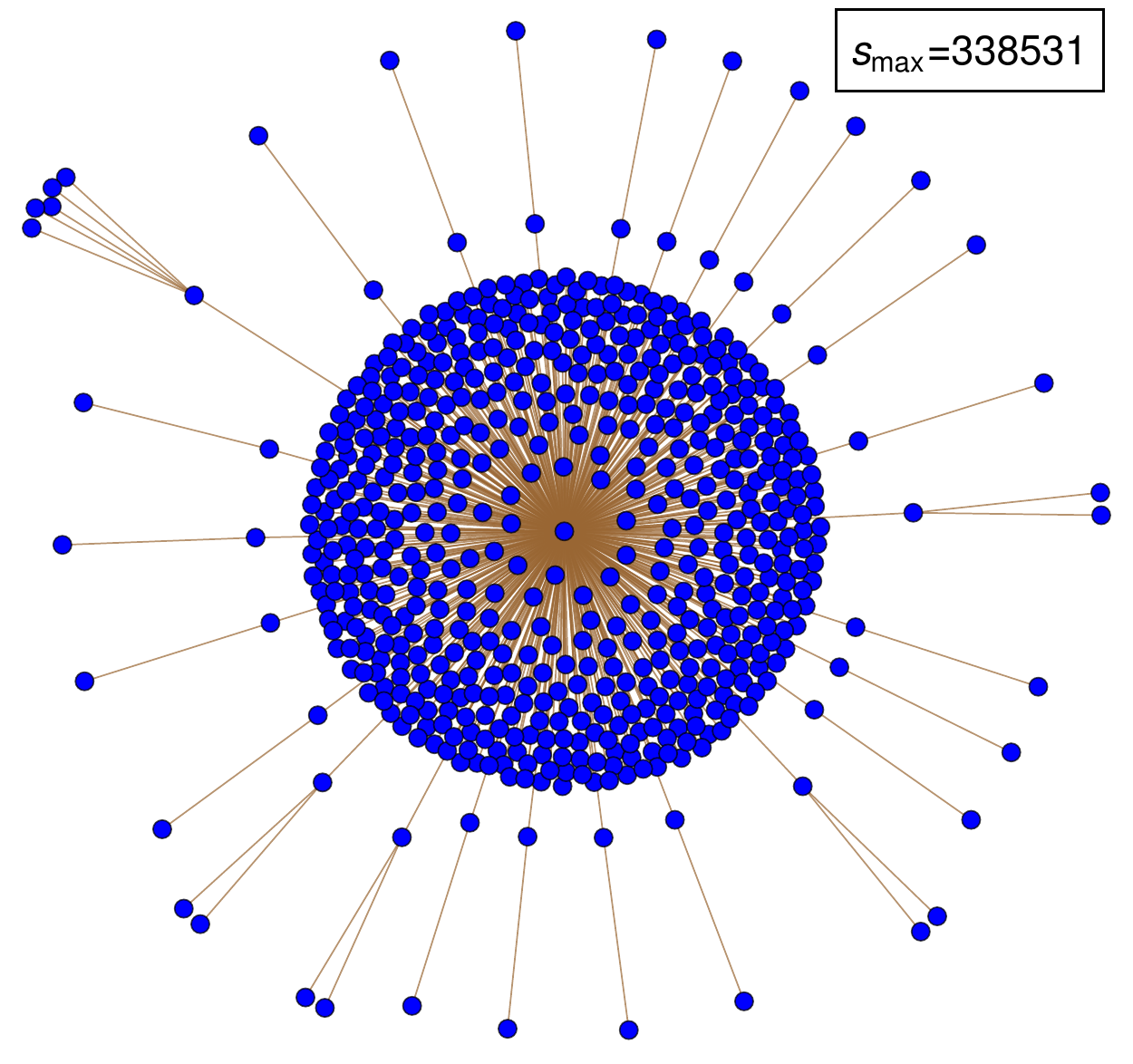}~~~~~~~~~
  \includegraphics[width=4.5cm]{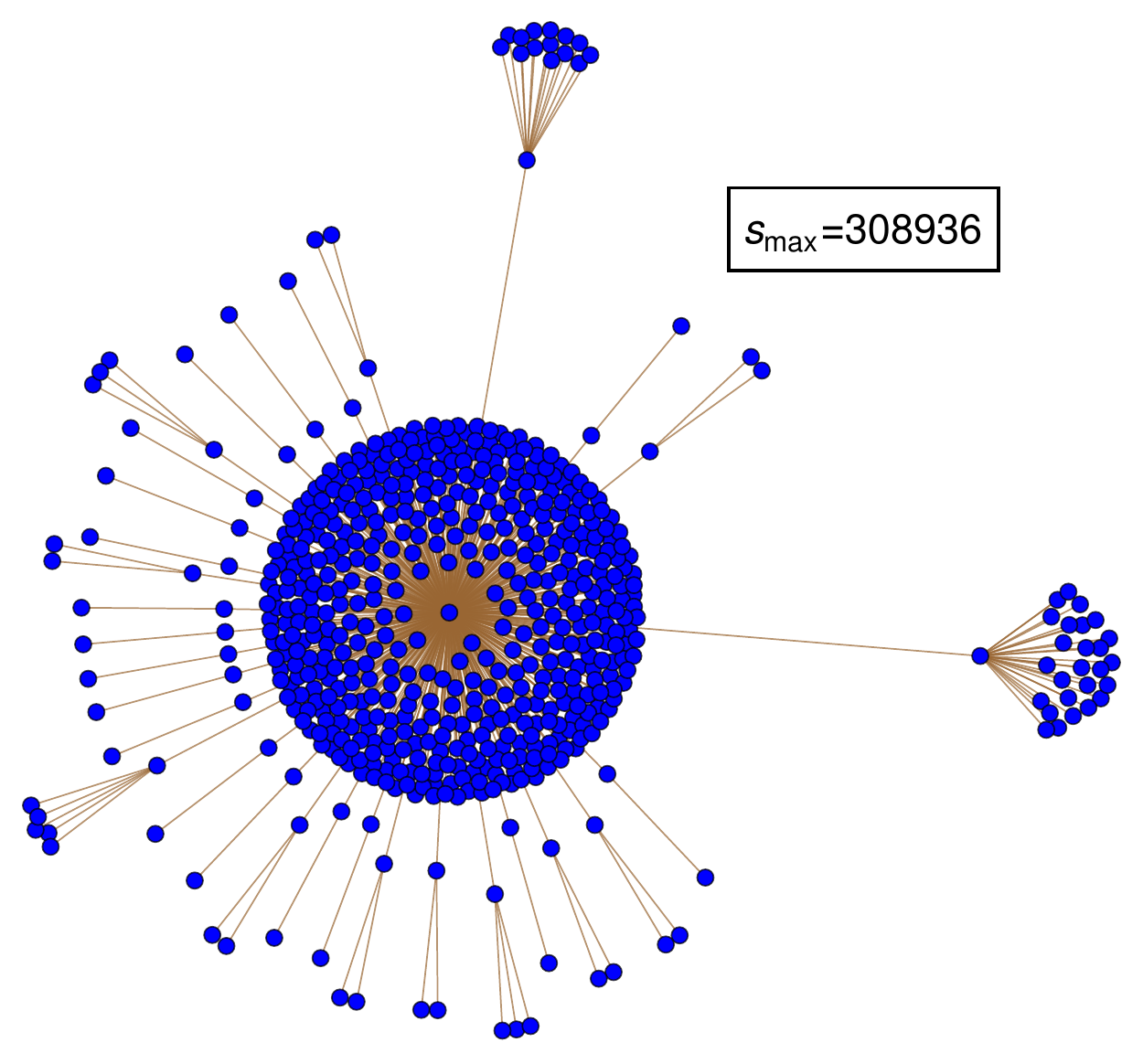}~~~~~~~~~
  \includegraphics[width=4.5cm]{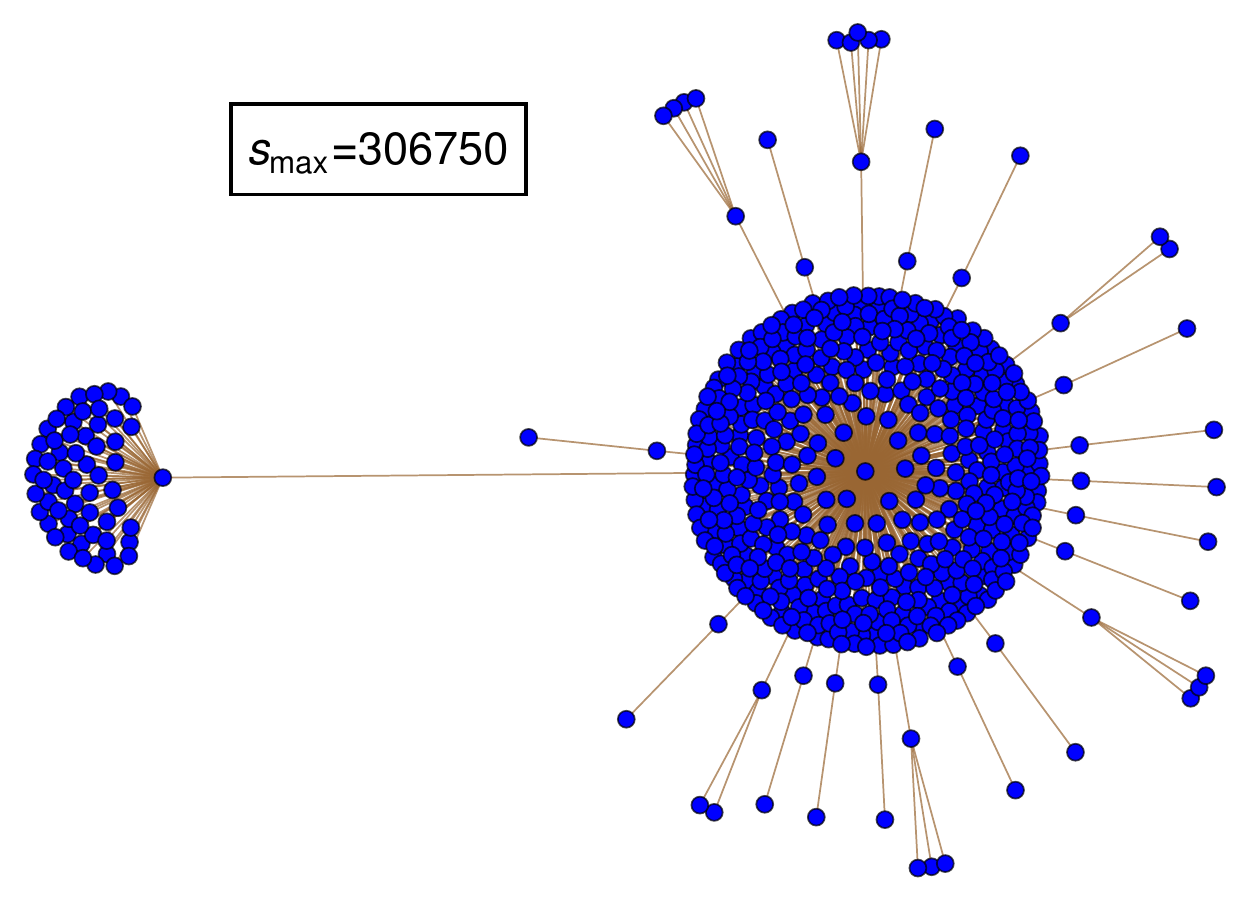}
  \caption{\label{fig:appmodBA} {\it Top}: representative graphs generated using the ModBA model kernel Eq.~\ref{eq:kernel2}, corresponding to m12i (left), m12r (middle), and m12z (right), where $N=600$. They are chosen from a sample of randomly-generated $100$ graphs, such that the graph distance is minimal with respective to those shown in Fig.~\ref{fig:self} (top), correspondingly. {\it Bottom}: reconstructed graphs with $s_G=s_{\rm max}$ based on the degree-preserving rewiring algorithm in Ref.~\cite{Li2005}. }
\end{figure*}

Fig.~\ref{fig:appmodBA} (top) shows representative graphs generated using the ModBA model, corresponding to m12i (left), m12r (middle), and m12z (right) in Fig.~\ref{fig:self}. These graphs are chosen in the following way. We first use the kernel in Eq.~\ref{eq:kernel2}, fix the node number to be $N=600$, and generate $100$ graphs randomly. For each FIRE2 graph, we then calculate its graph distance with the $100$ ModBA graphs $d_A=\sqrt{\sum^N_{i=1}(\lambda^{\rm FIRE2}_{i}-\lambda^{\rm ModBA}_{i})^2}$, where $\lambda^{\rm FIRE2}_{i}$ and $\lambda^{\rm ModBA}_{i}$ are the $i^{\rm th}$ eigenvalues of adjacency matrices for the FIRE2 and ModBA graphs, respectively, and we pick up the one that has the smallest distance to represent the FIRE2 system. Fig.~\ref{fig:appmodBA} (bottom) shows reconstructed ModBA graphs with $s_G=s_{\rm max}$, based on the degree-preserving rewiring algorithm in~\cite{Li2005}. Through these examples, we explicitly demonstrate that our ModBA model can reproduce the scale-free and self-similar properties of cosmological structure formation.

\section{Calculation of the graph metric $s_G$}
\label{app:sg}

\begin{figure*}[htbp]
  \centering
  \includegraphics[height=6cm]{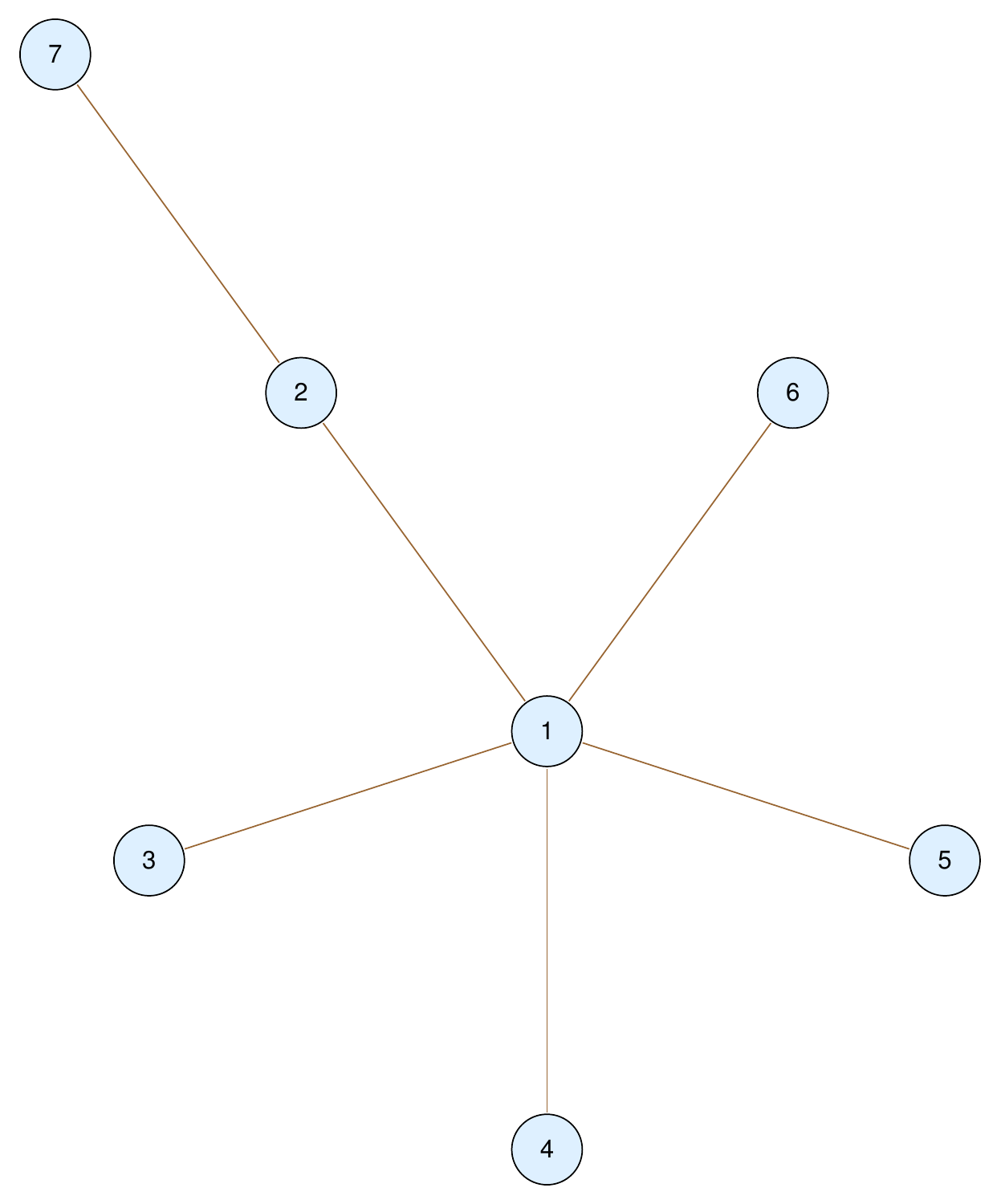}
  \caption{\label{fig:metric}An example graph that is used to demonstrate the calculation of the graph metric $s_G$. This is a part of the FIRE2 m12r system shown in Fig.~\ref{fig:cartoon}. }
\end{figure*}

We calculate the graph metric $s_G$ as $s_G=\sum_{(i,j)\in E}n_i n_j$, where $(i,j)$ denotes an edge connecting nodes $i$ and $j$, and $E$ denotes the set of all edges in a graph. Here we present an explicit calculation for a simple graph shown in Fig.~\ref{fig:metric}, which is a part of the FIRE2 m12r system, see Fig.~\ref{fig:cartoon}.  

Edge $(2,1)$: $n_2\times n_1=2\times5=10$; Edge $(3,1)$: $n_3\times n_1=1\times5=5$; Edge $(4,1)$: $n_4\times n_1=1\times5=5$; Edge $(5,1)$: $n_5\times n_1=1\times5=5$;  Edge $(6,1)$: $n_6\times n_1=1\times5=5$;  Edge $(7,2)$: $n_7\times n_2=1\times2=2$. Thus $s_G=\sum_{(i,j)\in E}n_i n_j=32$.

\end{document}